\documentclass[twocolumn,aps,prd,amssymb,eqsecnum,floatfix,nofootinbib]{revtex4}
\usepackage{epsfig}
\usepackage{amsmath}
\DeclareMathOperator{\erfc}{erfc}
\DeclareMathOperator{\erf}{erf}

\begin{document}

\title{Detection methods for non-Gaussian gravitational wave stochastic backgrounds}
\author{Steve Drasco\footnote{sd68@cornell.edu}}
%\email{sd68@cornell.edu}
\author{\'{E}anna \'{E}. Flanagan\footnote{eef3@cornell.edu.  Also
    Radcliffe Institute for Advanced Study, Putnam House, 10 Garden
    Street, Cambridge, MA 02138.}}
%\email{eef3@cornell.edu}
\affiliation{Newman Laboratory of Nuclear Studies, Cornell University, Ithaca, New York 14853}
%\affiliation{Center for Radiophysics and Space Research, Cornell University, Ithaca, New York 14853}
\date{\today}

\begin{abstract}
A gravitational wave stochastic background can be produced by a
collection of independent gravitational wave events.  There are two
classes of such backgrounds, one for which the ratio of the average
time between events to the average duration of an event is small
(i.e., many events are on at once), and one for which the ratio is
large.  In the first case the signal is continuous, sounds something
like a constant {\em hiss}, and has a Gaussian probability
distribution.  In the second case, the discontinuous or intermittent
signal sounds something like popcorn popping, and is described by a
non-Gaussian probability distribution.  In this paper we address the
issue of finding an optimal detection method for such a non-Gaussian
background.  As a first step, we examine the idealized situation in
which the event durations are short compared to the detector sampling
time, so that the time structure of the events cannot be resolved, and
we assume white, Gaussian noise in two collocated, aligned
detectors.  For this situation we derive an appropriate version of the
maximum likelihood detection statistic.  We compare the performance of
this statistic to that of the standard cross-correlation statistic
both analytically and with Monte Carlo simulations.  
In general the maximum likelihood statistic performs better than the
cross-correlation statistic when the stochastic background is
sufficiently non-Gaussian, resulting in a gain factor in the minimum
gravitational-wave energy density necessary for detection. 
This gain factor ranges roughly between 1 and 3, depending on the duty
cycle of the background, for realistic observing times and signal
strengths for both ground and space based detectors.
The computational cost of the statistic, although significantly greater 
than that of the cross-correlation statistic, is not unreasonable.
Before the statistic can be used in practice with real detector data,
further work is required to generalize our analysis to accommodate
separated, misaligned detectors with realistic, colored, non-Gaussian noise.
\end{abstract}



\maketitle

\section{Introduction and summary}
\label{s:Introduction and summary}

Along with a new generation of gravitational wave detectors around the world \cite{ligo,virgo,geo,tama}, detection 
algorithms for a variety of sources are nearing completion. If the
signals from these sources are  
detected, physicists stand to harvest unprecedented quantities of observational information concerning the 
nature of gravitation and the cosmos as a whole.  The fruit of this harvest will be the outputs of detection 
algorithms.  In this paper we introduce an algorithm designed for nearly optimal detection of a class of 
gravitational wave stochastic backgrounds. The non-Gaussian nature of this class of backgrounds causes 
the algorithm presented here to differ from the well studied cross-correlation based algorithms which are 
nearly optimal for Gaussian backgrounds.

\subsection{Gravitational wave stochastic backgrounds}
\label{ss:Gravitational stochastic backgrounds}

Consider a large collection of similar gravitational wave sources.  If
we cannot resolve the individual signals produced by these sources and
know only their statistical properties, the signals form a stochastic background.
A wide variety of candidate sources of gravitational wave stochastic backgrounds have been studied 
(for an excellent general review see Ref.~\cite{Allen Review}).
These include high redshift supernovae \cite{gaussian supernovae, non gaussian supernovae}, 
the first stars or so-called population III objects \cite{first stars}, 
rapidly rotating young neutron stars \cite{gaussian neutron stars 1, gaussian neutron stars 2}, 
early universe phase transitions and cosmic strings \cite{cosmic strings, bubbles}, 
inflation \cite{inflation}, 
and high redshift compact binaries \cite{binaries}.

Detecting a gravitational wave stochastic background produced by any one of these candidate sources could 
provide information on a variety of topics ranging from the evolution
of the star formation rate \cite{Coward} to the numbers and sizes of posited extra dimensions \cite{Hogan}.  
Because of this, stochastic backgrounds have long been thought to be among the most interesting 
possible types of gravitational radiation.

\subsection{Gaussian stochastic backgrounds}
\label{ss:Gaussian stochastic backgrounds}

In order to develop detection methods, it is traditionally assumed that the individual events making up a
background are uncorrelated and sufficiently frequent for the background to be Gaussian.  That is, it is
assumed that the conditions for applicability of the central limit theorem are satisfied. 

Unlike electromagnetic waves, gravitational waves cannot be screened from a detector. 
Using a single gravitational wave detector,  there is no practical way to distinguish between 
detector noise and a stochastic background of gravitational waves.  
As a consequence the sensitivity of a single detector to gravitational backgrounds is severely limited.
By comparing the outputs of multiple detectors, sensitivity levels can be enhanced.
Michelson \cite{Michelson} was the first to give a detailed description of such a detection 
method for a Gaussian stochastic background of gravitational waves in the presence of Gaussian 
detector noise.  His detection strategy and its later refinements \cite{Christensen,Flanagan,Allen Romano} are 
often referred to as the cross-correlation method.  
Recently the cross-correlation method has been modified to treat more realistic detectors
which themselves have sources of non-Gaussian noise \cite{robust
gaussian, robust gaussian II, Klimenko and Mitselmakher}.

We now briefly review the cross-correlation method.
Consider two gravitational wave detectors.   
The output of each detector is a collection of dimensionless strain measurements.  
Suppose that $N$ such measurements are made with each detector at regular time intervals. Denote these measurements 
by a $N \times 2$ matrix $h$ with components $h_i^k$,
where $i=1,2$ labels the detector, and $k=1,2,\ldots,N$ is a time index.  
To determine whether or not the data $h$ contains some desired signal,
one usually  
compares the value of some detection statistic $\Gamma(h)$ to a threshold value $\Gamma_*$.  That is, 
if $\Gamma(h) > \Gamma_*$ one concludes that a signal is present and otherwise
one concludes that no signal is present.
A detection statistic is said to be optimal if it yields the smallest probability of mistakenly concluding a signal 
is present (false alarm probability) after choosing a threshold which fixes the probability for 
mistakenly concluding a signal is absent (false dismissal probability).

Assume that the two detectors are collocated and aligned, and that each detector has white Gaussian noise with vanishing 
mean with no correlations between the two detectors. Then the standard cross-correlation detection 
statistic $\Lambda_{\text{CC}}$ for a Gaussian signal is 
\begin{equation} \label{cross correlation}
\Lambda_{\text{CC}}(h) = \frac{\hat\alpha^2 }{ \bar \sigma_1 \bar \sigma_2},
\end{equation}
where 
\begin{eqnarray}
\hat\alpha^2 &=& {\bar \alpha}^2 \theta({\bar \alpha}^2), \\
\bar\alpha^2 &=& \frac{1}{N}\sum_{k=1}^N h_1^{k} h_2^{k}, \\ 
\bar\sigma_i^2 &=& \frac{1}{N} \sum_{k=1}^N \left(h_i^k\right)^2, \label{intro bar sigma}
\end{eqnarray}
for $i=1,2$, and $\theta(x)$ is the Heaviside step function defined by
\begin{equation} 
\label{stepfunction}
\theta(x) = \left\{ 
\begin{array}{ll}
        1 & \text{ if } x \ge 0 \\
        0 & \text{ if } x  < 0
\end{array}
\right. .
\end{equation} 
This statistic is nearly optimal and can be derived 
from a maximum likelihood framework (see Sec.~\ref{ss:Gaussian
signal}). The subscript CC in $\Lambda_{\text{CC}}$
denotes ``cross correlation''.  The generalization of this statistic
to allow for colored noise and non-collocated, non-aligned detectors is
discussed in Refs.\ \cite{Michelson,Christensen,Flanagan,Allen Romano}.

% latex bug (?) causes start of next subsection to be squashed up
% against above text, the ~ below fixes is.

~

\subsection{Non-Gaussian stochastic backgrounds}
\label{ss:Non-Gaussian stochastic backgrounds}

A particular class of events will produce a Gaussian background
if, on average, at any given moment, many individual events are arriving at the detector. 
However, if the ratio of average time between events to the average duration
of events is large, then there are long stretches of ``silence'' or time during which no events arrive at
the detector.  The resulting stochastic background is non-Gaussian as the conditions for the applicability 
of the central limit theorem are not  satisfied. Recent work has suggested that some candidate 
gravitational wave stochastic backgrounds, of both cosmological and astrophysical origin, may  be 
non-Gaussian \cite{non gaussian supernovae, cosmic strings, first stars}.  However,
predictions concerning the properties of most gravitational wave background sources rely heavily on theoretical
arguments which extrapolate well beyond observational support.  Such extrapolations are always in some sense
speculative. It is conceivable that backgrounds predicted to be Gaussian may in fact turn out to be non-Gaussian, 
or vice versa.  

In Sec.~\ref{ss:Non-Gaussian signal} below, we apply a maximum
likelihood framework to derive a detection statistic for a particular
model of non-Gaussian stochastic background, which we now describe.
Let $h_i^k$ be the outputs of two collocated aligned gravitational
wave detectors with white, zero-mean, Gaussian noise with no
correlations between the two detectors.  The detector outputs $h_i^k$
consist of noise $n_i^k$ together with a common signal $s^k$:
\begin{eqnarray}
h_1^k &=& n_1^k + s^k \label{eq:common} \\
h_2^k &=& n_2^k + s^k. \nonumber
\end{eqnarray}
We wish to detect a non-Gaussian signal $s^k$ composed of long stretches of
silence which separate short bursts whose amplitudes are Gaussianly
distributed, and whose durations are smaller than the detector
resolution time (see Fig.~\ref{signal sketch}).  We therefore assume that each signal sample $s^k$ is
statistically independent with probability distribution [cf.\ Eq.\
(\ref{signal prior}) below]
\begin{equation}
p(s) = \xi {1 \over \sqrt{2 \pi} \alpha} \exp \left[-{s^2 \over 2
\alpha^2} \right] + (1 - \xi) \delta(s).
\label{eq:sigg}
\end{equation}
The parameter $\xi$ is what we call the {\it 
Gaussianity parameter} of the
stochastic background; it is the probability that, at any randomly
chosen time, a burst is present in the detector.  Thus $\xi$ takes
values in the 
interval $0 \le \xi \le 1$, and if $\xi=1$ then the background is
Gaussian.  The parameter $\xi$ can also be thought of as the duty
cycle of the background.  The parameter $\alpha$ in Eq.\ (\ref{eq:sigg}) is
the rms amplitude of the bursts.

Our nearly-optimal detection statistic
$\Lambda_{\text{ML}}^{\text{NG}}$ for the signal model (\ref{eq:sigg})
is given by [cf.\ Eq.\ (\ref{main result2}) below]
\begin{widetext}
\begin{eqnarray}  \label{main result}
\Lambda_{\text{ML}}^{\text{NG}}(h) &=&
\max_{0<\xi\le 1}~ \max_{\alpha > 0}~ \max_{\sigma_1 \ge 0}~ \max_{\sigma_2 \ge 0}~ \prod_{k=1}^N
\left\{  
        \frac{ \bar\sigma_1 \bar\sigma_2 \xi}{\sqrt{\sigma^2_1 \sigma^2_2 + \sigma^2_1 \alpha^2 + \sigma^2_2 \alpha^2}} 
        \exp \left[ \frac{\left( \frac{h_1^k}{\sigma^2_1} + \frac{h_2^k}{\sigma^2_2}\right)^2}
        {2\left( \frac{1}{\sigma^2_1} + \frac{1}{\sigma^2_2} + \frac{1}{\alpha^2} \right)} 
        - \frac{\left( h_1^k\right)^2}{2\sigma^2_1} - \frac{\left( h_2^k\right)^2}{2\sigma^2_2} + 1\right]  \right. \nonumber \\ 
&+& \left. \frac{\bar\sigma_1 \bar\sigma_2}{\sigma_1 \sigma_2}  (1-\xi)
        \exp \left[ - \frac{\left( h_1^k\right)^2}{2\sigma^2_1} - \frac{\left( h_2^k\right)^2}{2\sigma^2_2} + 1\right]\right\}.
\end{eqnarray}
\end{widetext}
Here the quantities $\bar\sigma_1$ and $\bar\sigma_2$ are defined by Eq.~(\ref{intro bar sigma}). 
The values of $\xi$, $\alpha^2$, $\sigma^2_1$ and $\sigma^2_2$ which achieve the maximum in Eq.~(\ref{main result}) are, respectively,  estimators of 
the signal's Gaussianity parameter, the variance of the signal events, and the variances of the noise in the two detectors.
If we calculate the quantity (\ref{main result}) at $\xi = 1$, instead of maximizing over $\xi$, the result is a statistic which is 
equivalent to the standard cross-correlation statistic
$\Lambda_{\text{CC}}$.

The subscript ML on $\Lambda_{\text{ML}}^{\text{NG}}$ stands for
``maximum likelihood'', while the superscript NG stands for
``non-Gaussian statistic''.  The superscript NG does {\it not}
necessarily mean that one is considering a non-Gaussian signal; both
of the statistics $\Lambda_{\rm CC}$ and 
$\Lambda_{\text{ML}}^{\text{NG}}$ can be applied to data containing
either a Gaussian signal or a non-Gaussian signal.

If the burst-amplitude parameter $\alpha$ is sufficiently large 
and the bursts are well separated in time, then the
individual bursts can 
be seen in the detector output.  In this 
case one could use, for example, the simple burst statistic
\footnote{In reality the statistic (\ref{eq:lambdaBdef}) would
be especially susceptible to non-Gaussian noise bursts in the detector
and so would not be used in practice; instead one would need search
for events where $|h_1^k|$ and $|h_2^k|$ are simultaneously large.  In
this paper we restrict attention for simplicity to Gaussian detector
noise; it will be important for future more general analyses to 
to allow for (uncorrelated) non-Gaussian noise components in the two
detectors.} 
\begin{equation}
\Lambda_\text{B} \equiv \max_{1 \le k \le N} \ \left| h_1^k \right|.
\label{eq:lambdaBdef}
\end{equation}
on the data from detector 1 to detect the signal.  The burst statistic
(\ref{eq:lambdaBdef}) and the cross-correlation statistic
$\Lambda_\text{CC}$ are used as references for comparison for 
the maximum likelihood statistic below.

\subsection{Main results}
\label{ss:Main results}

There are two main results in this paper.  The first result is the detection statistic $\Lambda_{\text{ML}}^{\text{NG}}$ given by Eq.~(\ref{main result}),
which is derived in Sec.~\ref{ss:Non-Gaussian signal}. This statistic is nearly optimal for 
the detection of a class of non-Gaussian gravitational wave stochastic backgrounds incident on a pair of 
idealized detectors.

\begin{figure}
\begin{center}
\epsfig{file=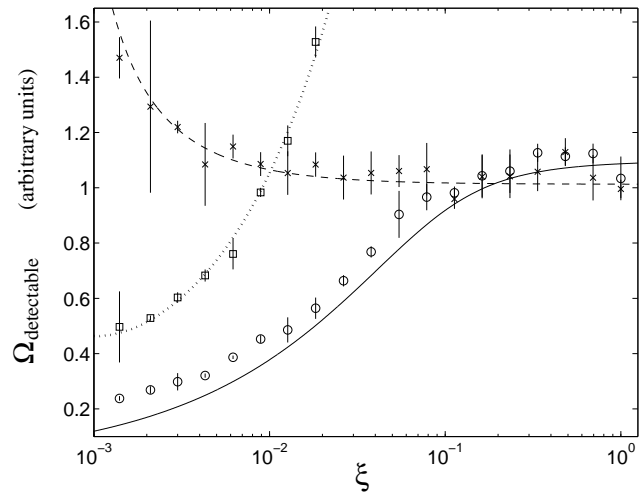,width=8.5cm}
\caption{
This plot shows the minimum gravitational-wave energy density
$\Omega_{\rm detectable}$ necessary for detection, for several
different detection statistics, as a function of the
background's Gaussianity parameter $\xi$.  
The Gaussianity parameter $\xi$ is the probability that, 
at any randomly chosen time, the waves from an event are incident on
the detectors, and thus takes values in the interval $0 \le \xi \le
1$.  For a Gaussian background $\xi=1$.
The circles are the results of our Monte Carlo simulations for the
maximum likelihood statistic $\Lambda_\text{ML}^\text{NG}$, and the 
solid curve shows the approximate theoretical prediction (\ref{eq:ansA}) and
(\ref{eq:ansB}) for 
this statistic (expected to be accurate only to within a few tens of
percent).   
The crosses are the Monte Carlo results for the
cross-correlation statistic $\Lambda_\text{CC}$, and the 
dashed curve shows the theoretical prediction \protect{(\ref{analytic
detectable})} for 
this statistic.  Finally the squares are the Monte Carlo results for the
burst statistic \protect{(\ref{eq:lambdaBdef})}, and the dotted curve shows the
corresponding theoretical prediction given by Eqs.\ \protect{(\ref{burstans1})}
and \protect{(\ref{burstans2})}.
For each statistic, the vertical error bars on the Monte Carlo
simulation results give the fluctuations computed from 4 different
runs, each with 2000 trials.  
The number of data points is $N = 10^4$, and the false alarm and false
dismissal 
probabilities are both $0.1$.  
A detailed description of the
simulations and the analytical predictions can be found in
Sec.~\ref{s:Performance comparison}.  
}
\label{omega gain}
\end{center}
\end{figure}

\begin{figure}
\begin{center}
\epsfig{file=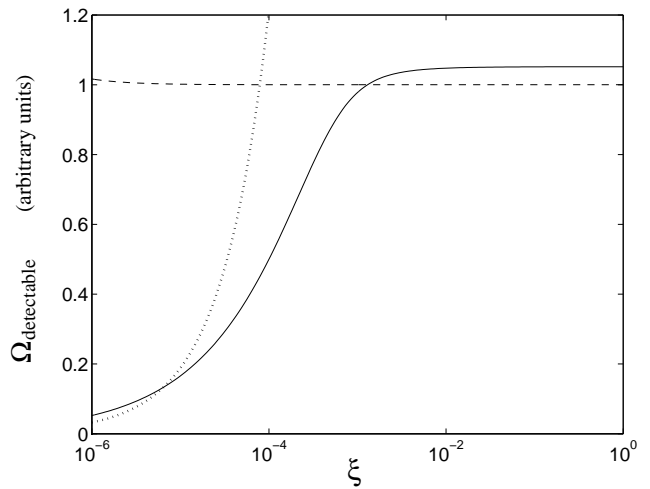,width=8.5cm}
\caption{
The minimum gravitational-wave energy density $\Omega_{\rm
detectable}$ necessary for detection as a function of the
background's Gaussianity parameter $\xi$ for $N = 10^9$ data points,
which is a realistic number of data points for ground based detectors.
The false alarm and false dismissal probabilities are both 0.01.
The solid line is the theoretical prediction (\ref{eq:ansA}) and
(\ref{eq:ansB}) for the maximum
likelihood statistic, which is expected to be accurate to a few tens
of  percent.  The dashed line is the theoretical prediction
(\protect{\ref{analytic}}) for the cross correlation statistic, and
the dotted line is the theoretical prediction
(\ref{burstans1})--(\ref{burstans2}) for the burst 
statistic; see caption to Fig.\ \protect{\ref{omega gain}}.
This plot indicates a maximum gain factor of $\sim 2$ in energy
density for duty cycles  
in a narrow band near $\xi \sim 10^{-4}$.}
\label{fig:theoretical}
\end{center}
\end{figure}

The second main result, summarized in Figs.~\ref{omega gain} and \ref{fig:theoretical}, is a
comparison of  
the performances of the maximum likelihood statistic
$\Lambda_{\text{ML}}^{\text{NG}}$, the  
cross-correlation statistic $\Lambda_{\text{CC}}$, and the burst
statistic $\Lambda_\text{B}$.  
That comparison is quantified in terms of the the minimum
gravitational-wave energy density  $\Omega_\text{detectable}$ 
necessary for detection.   The values of this quantity for the three
different statistics $\Lambda_\text{ML}^\text{NG}$,
$\Lambda_\text{CC}$ and $\Lambda_\text{B}$ we will denote by
$\Omega_\text{detectable}^\text{ML}$, $\Omega_{\rm
detectable}^\text{CC}$, and $\Omega_\text{detectable}^\text{B}$,
respectively.  Results for these three quantities obtained from Monte
Carlo simulations are shown in Fig.\ \ref{omega gain}, which gives
$\Omega_\text{detectable}$ as a function of $\xi$ for $N = 10^4$ data
points.  The Monte Carlo simulations are described in Sec.\
\ref{ss:Description of the simulation algorithm}
below.  The figure shows that in the limit $\xi \to 1$ of Gaussian
signals, the statistics $\Lambda_\text{ML}^\text{NG}$ and
$\Lambda_\text{CC}$ perform approximately equivalently (the
cross-correlation statistic is slightly better).  As the Gaussianity
parameter $\xi$ is decreased, the performance of
$\Lambda_\text{ML}^\text{NG}$ improves, until at $\xi \sim 10^{-2.5}$ it
is better than that of $\Lambda_\text{CC}$ by about a factor of $3$ in
energy density.  Finally, in the 
limit $\xi \to 0$, the individual bursts become visible and the burst
statistic $\Lambda_\text{B}$ becomes the best statistic.

{}Figure \ref{omega gain} also shows theoretical curves for the three
quantities $\Omega_\text{detectable}^\text{ML}$, $\Omega_{\rm
detectable}^\text{CC}$, and $\Omega_\text{detectable}^\text{B}$.
These curves are derived and discussed in Sec.\ \ref{s:Performance
comparison} below.  For the burst and cross-correlation statistics,
the theoretical curves should have a fractional accuracy $\sim
1/\sqrt{N}$. For the maximum likelihood statistic, the theoretical
prediction is expected to be accurate to a few tens of percent.  These
expected accuracies are confirmed by the Monte Carlo simulations, as
seen in Fig.\ \ref{omega gain}.

The value $N = 10^4$ of the number of data points is roughly
appropriate for a space based detector like LISA, for which the
duration of a measurement might be $\sim 1 $ year and the effective bandwidth
$\sim 10^{-3}$ Hz.  However, for year-long observations with 
ground based detectors, the effective bandwidth will be $\sim 100$ Hz
and consequently the appropriate value of $N$ is $ \sim 10^9$.  We were 
unable to perform Monte Carlo simulations for this large value of $N$ due to
limitations in available computing power.  However, we show in Fig.\
\ref{fig:theoretical} the theoretical curves for the three different
statistics as functions of $\xi$ for $N=10^9$.  In this case, the
maximum likelihood statistic starts to outperform the
cross-correlation statistic at $\xi \sim 10^{-3}$, and the maximum
gain factor in energy density is of order $\sim 2$.

We next discuss the computational cost of the maximum likelihood
statistic $\Lambda_{\rm ML}^{\rm NG}$.  As is well known, the
computational cost of trying to detect a stochastic background using
the cross-correlation statistic $\Lambda_{\text{CC}}$ is   
negligible when compared to, say, matched-filter-based inspiral
waveform searches.  However, because of the non-trivial maximization
in Eq.~(\ref{main result}), the maximum likelihood statistic  
$\Lambda_{\text{ML}}^{\text{NG}}$ 
is computationally intensive.  In fact, every evaluation of the function
to be maximized over the four parameters 
$\xi$, $\alpha$, $\sigma_1$, and $\sigma_2$ requires computing a
length-$N$ sum or product, where $N$ is the number of data points,
and takes longer than the
entire cross-correlation detection method.  Depending on the method of calculation,
the computational cost of computing $\Lambda_{\text{ML}}^{\text{NG}}$ is larger than that
of computing $\Lambda_{\text{CC}}$ by a factor anywhere from $10^2$ to $10^4$.

To summarize, under the idealized assumptions of this paper, if one
searches for a stochastic background using the standard
cross-correlation statistic, then one might not detect a signal that
would have been detectable using our maximum likelihood statistic.
This conclusion probably generalizes to realistic detector
noise models and detector orientations.

\subsection{Outline of this paper}
\label{ss:Outline of this paper}

In Sec.~\ref{s:General theory of detection statistics and parameters estimator} we introduce notation, 
review the general theory of signal detection and parameter measurement, and derive a general form of the maximum 
likelihood detection statistic.  Then, in Sec.~\ref{s:Application to stochastic background searches}, we derive the maximum likelihood 
statistics for both a Gaussian background (Sec.~\ref{ss:Gaussian
signal}) and for the model (\ref{eq:sigg}) of a non-Gaussian background 
(Sec.~\ref{ss:Non-Gaussian signal}), assuming two idealized detectors.  
In Sec.~\ref{s:Performance comparison} we discuss analytical calculations and Monte Carlo simulations 
comparing the performance of 
the maximum likelihood and cross-correlation detection statistics.
Also in Sec.~\ref{s:Performance comparison} we show how the signal
parameters $\xi$ and $\alpha$ can be estimated, with reasonable
accuracy, for a strong non-Gaussian background. We conclude in
Sec.~\ref{s:Conclusions} with a  
discussion of the results.  

\section{General theory of detection statistics and parameter estimation}
\label{s:General theory of detection statistics and parameters estimator}

In this section we review various formal aspects of the theory of
signal detection and measurement. 
We derive a form of the maximum likelihood detection statistic that is
more general than has been considered before in  
the context of gravitational wave data analysis \cite{Allen Romano, general method, excess power, sam joe, sam unpublished}.
The material in this section can be found in a variety of texts
\cite{maximum likelihood}; we include this section for completeness
and to introduce notation. 

\subsection{Notational conventions}
\label{ss:Notational conventions}

We use calligraphic letters $\mathcal{A, B, C, \ldots}$ to denote
random variables.
As described in Sec.~\ref{ss:Gaussian stochastic backgrounds}, given
$D$ detectors we can assemble an $N \times D$ detector 
output matrix $\mathcal{H}$ with components $\mathcal{H}_i^k$ where $k=1,2,\ldots,N$ is
a time index, and $i=1,2,\dots,D$ labels the detector.
We assume that the detector outputs are made up of noise $\mathcal{N}$ and signal $\mathcal{S}$ 
with components $\mathcal{N}_i^k$ and $\mathcal{S}_i^k$ respectively, such that
\begin{equation} \label{detector output matrices} 
\mathcal{H = N + S}.
\end{equation}
Specific realizations of random variables will be denoted by lower case
Roman symbols.  For example,  
$h=n+s$ is a specific realization of Eq.~(\ref{detector output
matrices}), where the components of $h$ are $h_i^k$.

Probability densities for random variables will always be denoted by a lowercase $p$ and will carry a subscript
to indicate which random variable is being described.  For example, $p_\mathcal{N}(n)d^{ND}n$
is the probability that $n<\mathcal{N}< n+dn$, where $d^{ND}n$ is the product
\begin{equation}  \label{differential product}
d^{ND}n = \prod_{k=1}^{N}\prod_{i=1}^{D}dn_i^k.
\end{equation}
We write the normalization requirement for $p_\mathcal{N}(n)$ as
\begin{equation} \label{normalization} 
1 = \int d^{ND}n~ p_\mathcal{N}(n).
\end{equation}
Unless otherwise specified, integrals are over $\mathbb{R}^{ND}$ where $\mathbb{R}$ is the set of real numbers.

We assume a detector noise model with $Q_n$ parameters.  Let
$\mathcal{V}_n$ be a vector of length $Q_n$   
whose components are the parameters characterizing the noise in the
detectors.  We denote by  
$\Theta_n$ the space of all possible values of $\mathcal{V}_n$. Here
the subscript $n$ is not an index; it is merely short for ``noise''. 
We denote joint probabilities in the usual way.  For example, $p_{{\mathcal N},{\mathcal V}_n}(n,{\bf v}_n)d^{ND}n~d^{Q_n}v_n$ is 
the probability that $n<\mathcal{N}< n+dn$ and ${\bf v}_n<\mathcal{V}_n< {\bf v}_n+d{\bf v}_n$, where $d^{Q_n}v_n$ is defined by
\begin{equation} 
d^{Q_n}v_n = \prod_{l=1}^{Q_n} dv_n^l,
\end{equation}
and $dv_n^l$ is the $l$th component of $d{\bf v}_n$.
We also use vertical bars to denote conditional 
probabilities.  For example
\begin{equation} \label{conditional joint}  
p_{\mathcal{N|V}_n}(n|{\bf v}_n) d^{ND}n =\frac{ p_{\mathcal{N,V}_n}(n,{\bf v}_n) d^{Q_n}v_n}{ p_{\mathcal{V}_n}({\bf v}_n) d^{Q_n}v_n}d^{ND}n
\end{equation}
is the probability that $n<\mathcal{N}< n+dn$ given that ${\bf
v}_n<\mathcal{V}_n< {\bf v}_n+d{\bf v}_n$. 

We will often use the so-called total probability theorem \cite{Papoulis} to write probability densities 
for a specific random variable as an integral over the functional dependencies of that random variable. 
An example is
\begin{equation} \label{total probability} 
p_\mathcal{N}(n) =\int_{\Theta_n}d^{Q_n}v_n~ p_{\mathcal{N|V}_n}(n|{\bf v}_n) p_{\mathcal{V}_n}({\bf v}_n).
\end{equation}
Expanding probability densities in this way allows us to treat
parameters, such as the noise parameters $\mathcal{V}_n$ in 
Eq.~(\ref{total probability}), as unknowns.  In fact, such a treatment of
the noise parameters
is the crucial difference between the derivations of this work and those in previous studies of 
gravitational wave data analysis techniques \cite{Allen Romano, general method, excess power, sam joe, sam unpublished}.

We assume that the signal model contains $Q_s$ parameters, which we
will treat as random variables  
like the noise parameters.  We will denote by ${\mathcal V}_s$ the
random vector of length $Q_s$ containing the signal parameters, 
and by $\Theta_s$ the space of all possible values of ${\mathcal V}_s$.

We define the notions of ``signal present'' and ``signal absent'' in terms of a partition of the space
$\Theta_s$ of signal parameters into a disjoint union
\begin{equation} 
\Theta_s = \Theta_{s0} \cup \Theta_{s1},
\end{equation} 
where $\Theta_{s0}$ corresponds to the signal being absent, and $\Theta_{s1}$ the signal being present.
We define the random variable $\mathcal{T}$, taking values $\mathcal{T}=0$ or $\mathcal{T}=1$, according to 
\begin{equation} 
\mathcal{T} = \left\{ 
\begin{array}{ll}
        1 & \text{ if } \mathcal{V}_s \in \Theta_{s1} \\
        0 & \text{ if } \mathcal{V}_s \in \Theta_{s0}
\end{array} \right. .
\end{equation}
Thus $\mathcal{T}=1$ corresponds to a signal being present, and
$\mathcal{T}=0$ to no signal being present.  We define
\begin{equation} 
p_{\mathcal{S|V}_s,\mathcal{T}}(s|{\bf v}_s,0) = \left\{
\begin{array}{ll}
        0  & \text{ if } {\bf v}_s \in \Theta_{s1} \\
        \delta^{ND}(s) & \text{ if } {\bf v}_s \in \Theta_{s0}
\end{array} \right. ,
\end{equation} 
where $\delta^{ND}(s)$ is the $N \times D$ dimensional Dirac delta function.
We denote by  $p_\mathcal{T,H}(t,h)d^{ND}h$ the probability that
$\mathcal{T}=t$ and that $h < \mathcal{H} < h+dh$, where $t=0$ or $1$.
Similarly 
\begin{equation} 
p_\mathcal{H|T}(h|t) d^{ND}h = \frac{ p_\mathcal{H,T}(h,t) }{ P_\mathcal{T}(t)}d^{ND}h
\end{equation} 
is the probability that $h < \mathcal{H} < h+dh$ given that $\mathcal{T}=t$.

We denote probabilities (as opposed to probability densities) with an uppercase $P$. For example $P_\mathcal{T}(1)$ is the probability that a signal 
is present, and $P_\mathcal{T}(0)$ is the probability that a signal is
absent.

Before examining the detector outputs, we may have some idea, say from previous experiments,  of the probability 
that a signal will be present. We denote this prior probability by $P^{(0)}$. We denote by $P^{(1)}$ the posterior 
probability that the signal is present after examining $\mathcal{H}$ in the context of all prior experiments etc.  
All posterior quantities have an implicit dependence on the detector outputs.  To simplify the notation 
we will not explicitly show this dependence.  For example, we write $P^{(1)}$ rather than the more cumbersome 
$P^{(1)}(\mathcal{H})$ for the posterior probability that a signal is present.

There are prior and posterior versions of all probability densities. When necessary we will append superscripts
of $(0)$ and $(1)$ to distinguish priors and posteriors respectively.
For example $p^{(1)}_{\mathcal{V}_n}({\bf v}_n) = p_{\mathcal{V}_n|\mathcal{H}}({\bf v}_n|h)$ is the posterior probability density for
$\mathcal{V}_n$. The posterior distribution for the noise can be expanded in terms of $p^{(1)}_{\mathcal{V}_n}({\bf v}_n)$ as
\begin{equation} 
p^{(1)}_\mathcal{N}(n) =\int_{\Theta_n}d^{Q_n}v_n~ p_{\mathcal{N|V}_n}(n|{\bf v}_n) p^{(1)}_{\mathcal{V}_n}({\bf v}_n).  
\end{equation} 

%%%%%%%%%%%%%%%%%%%%%%%%%%%%%%%%%%%%%%%%%%%%%%%%%%%%%%%%%%%%%%%%%%%%%%%%%%%%%%%%%%%%%%%%%%%%%%%%%%%%%%%

The conventions and symbols which have been introduced above are summarized in tables 
\ref{conventions} and \ref{symbols} respectively.  

\begin{table*}
\caption{\label{conventions} A summary of conventions introduced in Sec.~\ref{ss:Notational conventions}.}
\begin{ruledtabular}
\begin{tabular}{p{8.5cm}p{8.5cm}}
Convention & Example \\ \hline

Random variables are denoted by upper case calligraphic letters. & 
The detector output matrix is denoted by $\mathcal{H}$. \\

Specific realizations of random variables are denoted by lower 
case Roman letters. (see next convention)& 
A specific observation run may result in a specific detector output 
matrix $h$ or say $x$. These results would be denoted $\mathcal{H}=h$ 
and $\mathcal{H}=x$ respectively. \\ 

A lower case $p$ denotes a probability density function (PDF). 
It's subscript determines the quantities with which it is associated. &
The PDF for the detector output $\mathcal{H}$ as a function of $h$, or say $x$, 
is denoted by $p_\mathcal{H}(h)$ and $p_\mathcal{H}(x)$ respectively. \\

A comma in a PDF subscript and argument indicates a joint PDF. &
The joint PDF for $\mathcal{N}$ and $\mathcal{V}_n$ as a function of 
$n$ and ${\bf v}_n$ respectively is denoted by $p_{\mathcal{N},\mathcal{V}_n}(n,{\bf v}_n)$. \\

A vertical bar in a PDF subscript and argument indicates a conditional PDF. &
The conditional PDF for $\mathcal{N}$ and $\mathcal{V}_n$ as a function of
$n$ and ${\bf v}_n$ respectively is denoted by $p_{\mathcal{N}|\mathcal{V}_n}(n|{\bf v}_n)$. \\

An upper case $P$ denotes a probability. & 

The probability that $\mathcal{T}=1$ is denoted by $P_\mathcal{T}(1)$. \\

Prior and posterior quantities are denoted by superscripts of $(0)$ and $(1)$ respectively. &

The prior probability that a signal is present is denoted by $P^{(0)}$, while the posterior 
probability that a signal is present, after an observation $\mathcal{H}=h$, is denoted by 
$P^{(1)} = P_{\mathcal{T}|\mathcal{H}}(1|h)$.
\end{tabular}
\end{ruledtabular}
\end{table*}

\begin{table}
\caption{\label{symbols} A summary of symbols introduced in Sec.~\ref{ss:Notational conventions}.}
\begin{ruledtabular}
\begin{tabular}{cp{7cm}}
Symbol & {Meaning} \\ \hline
$\mathcal{H},h$ & detector output matrix \\
$\mathcal{N},n$ & noise contribution to detector output matrix \\
$\mathcal{S},s$ & signal contribution to detector output matrix \\
$N$ & number of strain samples taken from one detector  \\
$D$ & number of detectors \\
$Q_n$ & number of parameters in the model noise PDF \\
$Q_s$ & number of parameters in the model signal PDF \\
$\mathcal{V}_n,{\bf v}_n$ & the parameters of the model noise PDF \\
$\mathcal{V}_s,{\bf v}_s$ & the parameters of the model signal PDF \\
$\Theta_n$ & the space of all possible values of $\mathcal{V}_n$\\
$\Theta_s$ & the space of all possible values of $\mathcal{V}_s$\\
$\Theta_{s0}$ & the subspace of $\Theta_s$ for which a signal is absent\\
$\Theta_{s1}$ & the subspace of $\Theta_s$ for which a signal is present \\
$\mathcal{T},t$ & 1 if a signal is present ($\mathcal{V}_s \in \Theta_{s1}$), otherwise 0\\
$P^{(0)}$ & prior probability that a signal is present\\
$P^{(1)}$ & posterior probability that a signal is present\\
\end{tabular}
\end{ruledtabular}
\end{table}
%%%%%%%%%%%%%%%%%%%%%%%%%%%%%%%%%%%%%%%%%%%%%%%%%%%%%%%%%%%%%%%%%%%%%%%%%%%%%%%%%%%%%%%%%%%%%%%%%%%%%%%%

\subsection{Detection statistics}
\label{ss:Detection statistics}

To detect a signal one uses a detection statistic, say $\Gamma=\Gamma(\mathcal{H})$, that is some function
of the detector outputs ${\cal H}$.  A signal is said to have been
detected when 
$\Gamma$ exceeds some threshold value $\Gamma_*$.  

Denote by $P_\text{FD}(\Gamma_*)$ the probability of false dismissal, that is, the probability
that we fail to detect a signal which is actually present.  Similarly, let $P_\text{FA}(\Gamma_*)$ be
the probability that we claim to have detected a signal which in fact is absent---the probability of false alarm.
For given signal and noise models and for a given statistic $\Gamma$, the
false alarm and false dismissal probabilities generate a curve in the
$P_\text{FA}$-$P_\text{FD}$ plane parametrized by the threshold $\Gamma_*$.
Such curves depend on the number of detectors $D$, the number of data points $N$, 
the signal parameters $\mathcal{V}_s$, and the noise parameters $\mathcal{V}_n$.

\begin{figure}
\begin{center}
\epsfig{file=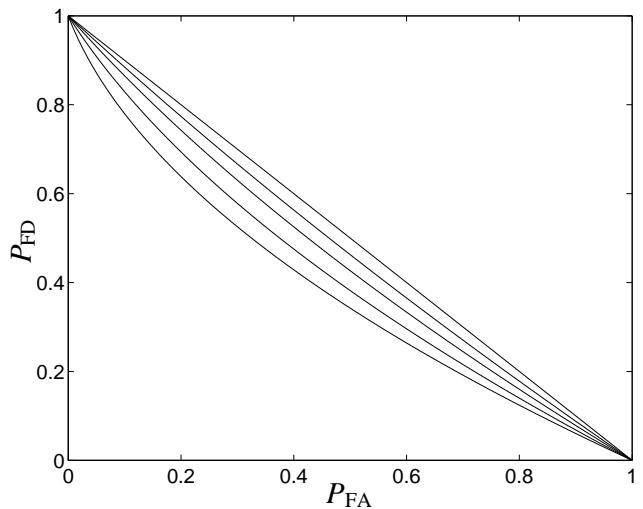,width=8.5cm}
\caption{False dismissal versus false alarm curves for typical detection statistics.}
\label{expected plots}
\end{center}
\end{figure}

Suppose that the statistic $\Gamma$ is bounded in the sense that 
there exist numbers $\Gamma_{\min}$ and $\Gamma_{\max}$ such that
$\Gamma_{\min} < \Gamma < \Gamma_{\max}$ for all ${\cal H}$. 
Then it is clear that $P_\text{FD}(\Gamma_{\min}) = 0$ and that
$P_\text{FA}(\Gamma_{\min})=1$.  As the threshold $\Gamma_*$
increases toward $\Gamma_{\max}$,  $P_\text{FD}(\Gamma_*)$ will
increase while $P_\text{FA}(\Gamma_*)$ 
decreases, until finally at $\Gamma_* = \Gamma_{\rm max}$, $P_\text{FD} = 1$, 
and $P_\text{FA} = 0$.  Thus, false dismissal-false alarm curves generally look 
something like those sketched in Fig.~\ref{expected plots}.

Note that if one uses a different statistic $f(\Gamma)$, where $f$ is
any function, then the shape of the $P_\text{FA}$-$P_\text{FD}$ curve does
not change as long as $f$ is monotonic in the sense that 
\begin{equation} \label{transformation}
\Gamma > \Gamma_* ~\Rightarrow~ f(\Gamma) > f(\Gamma_*).
\end{equation}
Only the parametrization of the curve changes under such a
transformation.  Statistics related by transformations $f$ satisfying
the monotonicity property (\ref{transformation}) have identical false
dismissal versus false alarm curves.

In 1933 Neyman and Pearson considered a simple signal detection scenario 
where the sets $\Theta_n$, $\Theta_{s1}$, and $\Theta_{s0}$ each contain a single element \cite{Neyman and Pearson}.  
They showed that for this scenario the detection statistic which minimizes $P_\text{FD}$ for any $P_\text{FA}$ 
is the so-called \emph{likelihood ratio} $\Lambda$, defined by
\begin{equation} \label{def1}
\Lambda = \frac{p_\mathcal{H|T}(h|1)}{p_\mathcal{H|T}(h|0)}.
\end{equation}
One notion of optimality for detection statistics is that the
statistic should minimize the false dismissal probability
at a fixed value of the false alarm probability.  For the simple
scenario above, this criteria, known as the  
Neyman-Pearson criteria, uniquely determines the likelihood ratio as the optimal statistic
\cite{Ferguson}.  However in general, when any of $\Theta_n$, $\Theta_{s1}$, or $\Theta_{s0}$ contains more than
one element, the statistic selected by this criteria is a function of the unknown parameters $\mathcal{V}_s$
and $\mathcal{V}_n$.  Thus, as is well known, the Neyman-Pearson
criteria does not single out a unique statistic in such cases.

In this paper we will obtain our detection statistics from Bayesian
considerations, but we will quantify their  
effectiveness using the Neyman and Pearson criteria of comparing false
dismissal probabilities at fixed false alarm probabilities.

\subsection{Likelihood ratio and likelihood function}
\label{ss:Likelhood ratio and likelihood function}

{}From a Bayesian point of view, a natural criterion for 
deciding that a signal is present  
is for the posterior probability $P^{(1)}$ to
exceed some threshold \cite{Bayes}. 
The posterior probability $P^{(1)}$ is related to the prior
probability $P^{(0)}$ and to the likelihood ratio $\Lambda$ defined by
Eq.~(\ref{def1}) by 
\begin{equation} \label{def2} 
\frac{P^{(1)}}{1-P^{(1)}} = \Lambda \frac{P^{(0)}}{1-P^{(0)}}.
\end{equation}
See appendix \ref{s:appendixA} for a derivation of Eq.~(\ref{def2}) in the most general context where the sets $\Theta_n$, 
$\Theta_{s1}$, and $\Theta_{s0}$ are all non-trivial. It follows from Eq.~(\ref{def2}) that $P^{(1)}$ is a monotonic function
of $\Lambda$, so thresholding on $P^{(1)}$ is equivalent to thresholding on $\Lambda$.
This makes $\Lambda$, or approximate versions of it, the natural choice for a detection statistic.

We derive in Appendix \ref{s:appendixA} the following general formula
for the likelihood ratio as a function of the data ${\cal H} = h$:
\begin{widetext}
\begin{equation} \label{general likelihood ratio}
\Lambda = \frac{\displaystyle  \int_{\Theta_{s1}} d^{Q_s}v_s~ \int d^{ND}s~ \int_{\Theta_n} d^{Q_n}v_n~ p_{\mathcal{N|V}_n}(h-s|{\bf v}_n) p_{\mathcal{V}_n}({\bf v}_n) 
                 p_{\mathcal{S|V}_s,\mathcal{T}}(s|{\bf v}_s,1) p_{\mathcal{V}_s|\mathcal{T}}({\bf v}_s|1) }
               {\displaystyle                       \int_{\Theta_n} d^{Q_n}v_n'~             p_{\mathcal{N|V}_n}(h|{\bf v}_n')  p_{\mathcal{V}_n}({\bf v}_n') }.
\end{equation}
\end{widetext}
The various probability distributions that appear in Eq.\ (\ref{general
likelihood ratio}) are (i) the prior distribution
$p_{\mathcal{V}_s|\mathcal{T}}({\bf v}_s|1)$ for the signal
parameters ${\bf v}_s$; (ii) the distribution 
$p_{\mathcal{S|V}_s,\mathcal{T}}(s|{\bf v}_s,1)$ for the signal $s$
given the signal parameters ${\bf v}_s$; (iii) the prior distribution 
$p_{\mathcal{V}_n}({\bf v}_n)$ for the noise parameters ${\bf v}_n$;
and (iv) the distribution $p_{\mathcal{N|V}_n}(h|{\bf v}_n)$ for the
noise $n$ given the noise parameters ${\bf v}_n$.

We can interpret Eq.\ (\ref{general likelihood ratio}) as follows.  
In the simple signal detection scenario, we choose between a pair of
simple claims: 
(i) $\mathcal{V}_s = {\bf v}_{s0}$ or (ii) $\mathcal{V}_s = {\bf v}_{s1}$.
In general we choose between a pair of complicated, or composite, claims:
(i)  $\mathcal{V}_s \in \Theta_{s0}$ or (ii) $\mathcal{V}_s \in
\Theta_{s1}$, where both $\Theta_{s0}$  
and $\Theta_{s1}$ contain many elements.
Equation (\ref{general likelihood ratio}) says that the best way to
chose between a pair of complicated claims is 
to 
first break the complicated pair of claims into pairs of simple
claims, then compute the likelihood ratio for each pair of simple claims,  
and sum the results of each choice. That is, the likelihood ratio can
be written as an integral over the parameters of the composite claims
\begin{equation} \label{likelihood function def}
\Lambda = \int_{\Theta_{s1}} d^{Q_s}v_s~ \int_{\Theta_n} d^{Q_n}v_n ~\Lambda({\bf v}_s,{\bf v}_n),
\end{equation} 
where the integrand $\Lambda({\bf v}_s,{\bf v}_n)$, which we refer to
as the \emph{likelihood function}, 
can be read off from Eq.\ (\ref{general likelihood ratio}):
\begin{widetext}
\begin{equation} \label{likelihood function} 
\Lambda({\bf v}_s,{\bf v}_n) = \frac{\displaystyle \int d^{ND}s~  p_{\mathcal{N|V}_n}(h-s|{\bf v}_n) p_{\mathcal{S|V}_s,\mathcal{T}}(s|{\bf v}_s,1) 
                                         p_{\mathcal{V}_n}({\bf v}_n) p_{\mathcal{V}_s|\mathcal{T}}({\bf v}_s,1) }
                    {\displaystyle \int_{\Theta_n} d^{Q_n}v_n'~ p_{\mathcal{N|V}_n}(h|{\bf v}_n')  p_{\mathcal{V}_n}({\bf v}_n') }.
\end{equation} 
\end{widetext}

The likelihood function\footnote{There are two different conventions for the definition of the likelihood function.
Some authors include the probability distributions for $\mathcal{V}_s$ and $\mathcal{V}_n$ in the definition
of $\Lambda({\bf v}_s,{\bf v}_n)$ as we have in Eq.~(\ref{likelihood function}), while others leave these out 
of $\Lambda({\bf v}_s,{\bf v}_n)$ and would show these distributions explicitly in 
Eq.~(\ref{likelihood function def}).}
%%%%%%%%%%%%%%%%%%%%%%%%%%%%%%%%%%%%%%%%%%%%%%%%%%%%%%%%%%%%%%%%%%%%%%%%%%%%%%%%%%%%%%%%%%%%%%%%%%%%%%%%%%%%%
$\Lambda({\bf v}_s,{\bf v}_n)$ can be used to
compute the posterior probability density 
$p^{(1)}_{\mathcal{V}_s,\mathcal{V}_n|\mathcal{T}}({\bf v}_s,{\bf
v}_n|1)$ for the signal and noise parameters given that a signal is
present, via the formula
\begin{equation} \label{distribution relation} 
\frac{ P^{(1)} }{ 1 - P^{(1)} } p^{(1)}_{\mathcal{V}_s,\mathcal{V}_n|\mathcal{T}}({\bf v}_s,{\bf v}_n|1)
= \Lambda({\bf v}_s,{\bf v}_n) \frac{ P^{(0)} }{ 1 - P^{(0)} }.
\end{equation}
A derivation of Eq.~(\ref{distribution relation}) can be found in appendix \ref{s:appendixA}.

\subsection{Maximum likelihood detection statistics and parameter estimators}
\label{ss:Maximum likelihood detection statistics and parameter estimators}

In many applications, it is impractical to compute the detection
statistic (\ref{general likelihood ratio}) because of the
multi-dimensional integrals involved \cite{Loredo}.  However,
approximate versions of the statistic are often easier to compute and
useful.  If a signal is present with sufficiently large amplitude, then
the integrand in the numerator of Eq.\ (\ref{general likelihood ratio})
will be sharply peaked.  The integrand in the denominator of 
Eq.\ (\ref{general likelihood ratio}) will also be sharply peaked when
there is sufficient data that the noise is well characterized.  Under
these circumstances, the integrals can be written as the values of the
corresponding integrands at the peaks multiplied by ``width
factors'', where the width factors depend only weakly
on the data $h$ and can be neglected without affecting much the
performance of the statistic.  [The width factors from the integrals
over the noise parameters will tend to cancel between the numerator
and denominator].  Also, frequently the prior distributions for
$\mathcal{V}_s$ and $\mathcal{V}_n$ are slowly varying, and neglecting
those distributions 
has a negligible effect on the performance of the statistic.   
Under these conditions the maximum likelihood detection statistic
$\Lambda_\text{ML}$ defined by  
\begin{widetext}
\begin{equation} \label{general likelihood estimator}
\Lambda_\text{ML} = \frac{ \displaystyle \max_{{\bf v}_s\in\Theta_{s1}}~\max_{{\bf v}_n\in\Theta_n}~ \int d^{ND}s ~p_{\mathcal{N|V}_n}(h-s|{\bf v}_n) 
			   p_{\mathcal{S|V}_s,\mathcal{T}}(s|{\bf v}_s,1) }
                         { \displaystyle \max_{{\bf v'}\in\Theta_n}~ p_{\mathcal{N|V}_n}(h|{\bf v}_n') }
\end{equation}
\end{widetext}
is a natural approximate version of $\Lambda$ 
% FOOTNOTE %%%%%%%%%%%%%%%%%%%%%%%%%%%%%%%%%%%%%%%%%%%%%%%%%%%%%%%%%%%%%%%%%%%%%%%%%%%%%%%%%%%%%%%%%%%%%%%%%%
\footnote{In the event that the priors for $\mathcal{V}_s$
and $\mathcal{V}_n$ restrict these parameters to regions $\Theta_{s1}' \subset \Theta_{s1}$ and
$\Theta_n' \subset \Theta_n$, the bounds of the maximizations in Eq.~(\ref{general likelihood estimator})
should be changed to $\Theta_{s1} \rightarrow \Theta_{s1}'$ and $\Theta_n \rightarrow \Theta_n'$.}. 
%%%%%%%%%%%%%%%%%%%%%%%%%%%%%%%%%%%%%%%%%%%%%%%%%%%%%%%%%%%%%%%%%%%%%%%%%%%%%%%%%%%%%%%%%%%%%%%%%%%%%%%%%%%%%
The subscript ML denotes that (\ref{general likelihood estimator}) is the maximum likelihood approximate 
version of $\Lambda$.
See Ref.~\cite{maximum likelihood} for further discussion of
$\Lambda_\text{ML}$ as an approximate version of $\Lambda$ \footnote{Note that $\Lambda_{\rm ML}$ is an 
approximate version of $\Lambda$ only in the sense that the false dismissal versus false alarm curves
of the two statistics will be close to one another.  The numerical
values of $\Lambda_{\rm ML}$ and $\Lambda$ will in general differ
significantly, due to the width factors and priors.  Therefore the
statistic $\Lambda_{\rm ML}$ cannot be used in Eq.\ (\ref{def2}) to
compute Bayesian thresholds for detection given a desired value of
$P^{(1)}$.}.

A particular special case of the detection
statistic (\ref{general likelihood estimator}), which is widely used,
is the following.  Assume that the noise   
parameters have some known values $\mathcal{V}_n = {\bf v}_n$. Then the noise priors and the $\Theta_n$ integrals 
in Eq.~(\ref{general likelihood ratio}) are trivial, and one obtains
the detection statistic
\begin{equation} \label{known likelihood estimator}
\tilde\Lambda_\text{ML}=\frac{\displaystyle \max_{{\bf v}_s\in\Theta_{s1}}~ \int d^{ND}s ~p_{\mathcal{N|V}_n}(h-s|{\bf v}_n) p_{\mathcal{S|V}_s,\mathcal{T}}(s|{\bf v}_s,1)}
			     {\displaystyle p_{\mathcal{N|V}_n}(h|{\bf v}_n)}.
\end{equation}
See Ref.~\cite{sam joe} for an exploration of the statistic (\ref{known likelihood estimator}) in the 
context of stochastic backgrounds.  We will show below that for a Gaussian stochastic background, 
$\Lambda_\text{ML}$ reduces to the standard cross-correlation
statistic while the more specialized statistic
$\tilde\Lambda_\text{ML}$ does not.  Thus for stochastic backgrounds,
treating the noise parameters as unknowns is crucial \cite{robust
gaussian II}.

When the noise and signal parameters $\mathcal{V}_n$ and
$\mathcal{V}_s$ can take on many values, one naturally would like to
know which  
values are realized. Equation (\ref{distribution relation}) suggests
using the values $\hat {\bf v}_n$ and $\hat {\bf v}_s$ defined by
\begin{equation} \label{ML estimates}
\Lambda(\hat {\bf v}_s, \hat {\bf v}_n) = \max_{{\bf v}_s\in\Theta_{s1}}~ \max_{{\bf v}_n\in\Theta_n}~ \Lambda({\bf v}_s,{\bf v}_n).
\end{equation}
The estimators $\hat {\bf v}_n$ and $\hat {\bf v}_s$ are known as maximum likelihood estimators. 
Note that ${\bf v}_s=\hat {\bf v}_s$ and ${\bf v}_n=\hat {\bf v}_n$ also maximize the numerator in Eq.~(\ref{general likelihood estimator}).
For the remainder of this paper we will use $\Lambda_\text{ML}$, defined by Eq.~(\ref{general likelihood estimator}), as our 
detection statistic, and $\hat {\bf v}_s$ and $\hat {\bf v}_n$, defined by Eq.~(\ref{ML estimates}), as parameter estimators.

\section{Application to stochastic background searches}
\label{s:Application to stochastic background searches}

In this section we derive the maximum likelihood detection statistic (\ref{general likelihood estimator})
for a simplified model of the detection problem for stochastic gravitational waves, and for a specific simple model 
of a non-Gaussian stochastic background.

\subsection{Assumptions}
\label{ss:Assumptions}

We assume two detectors with outputs $\mathcal{H}_i^k$, where $i=1,2$
labels the detector
and $k=1,2,\ldots,N$ is a time index. 
We assume that the noise in detector one is uncorrelated with the
noise in detector two.
We will require the noise in both detectors to have vanishing mean and to be both Gaussian and white, so that
\begin{equation} \label{assumption2}
p_{\mathcal{N|V}_n}\left[n|(\sigma_1,\sigma_2)\right] = 
\prod_{k=1}^N \frac{ 1 }{ 2 \pi \sigma_1\sigma_2 } \exp\left[- \frac{ (n_1^k)^2 }{ 2 \sigma^2_1 } - \frac{ (n_2^k)^2 }{ 2 \sigma^2_2 } \right].
\end{equation}
The parameters $\sigma_1$ and $\sigma_2$ in Eq.~(\ref{assumption2})
are the square roots of the variances of the noise 
in the two detectors. 
For this model ${\bf v}_n = (\sigma_1,\sigma_2)$ and $\Theta_n =
\left\{ (\sigma_1,\sigma_2) ~|~ \sigma_1 \ge 0 \text{ and } \sigma_2
\ge 0\right\}$. 

We assume that the detectors are collocated and aligned, so that the
same signal is present in both detectors
\begin{equation}
\mathcal{S}_1^k = \mathcal{S}_2^k = \mathcal{S}^k.
\end{equation}
Lastly we assume that the individual signal samples are uncorrelated
and identically distributed, i.e., the signal is white, so that
\begin{equation} \label{assumption4}
p_\mathcal{S}(s) = \prod_{k=1}^N p_{\mathcal{S}^k}(s^k).
\end{equation} 
Our assumptions (\ref{assumption2})-(\ref{assumption4})
are unrealistic for both ground-based and space-based detectors: we
expect the noise to be colored with significant non-Guasssian
components, and in general detectors will not be co-located and
aligned.  Our analysis is therefore just a first step, and will need
to be generalized.  However, we expect that our central conclusion ---
the existence of statistics which outperform the standard
cross-correlation statistic for nonGaussian signals --- is robust, and
will not be altered when these complications are taken into account.

We now derive a general formula for the maximum likelihood statistic 
(\ref{general likelihood estimator}), which we apply in both the
Gaussian and non-Gaussian cases in the following two subsections. 
The denominator in Eq.~(\ref{general likelihood estimator}) can be
written, from Eq.~(\ref{assumption2}), as
\begin{equation} \label{denominator to maximize} 
\max_{\sigma_1 \ge 0}~ \max_{\sigma_2 \ge 0}~ \left\{  \left(2 \pi \sigma_1\sigma_2 \right)^{-N} 
	                         	\exp\left[ -\frac{N}{2} \left( \frac{\bar\sigma^2_1}{\sigma^2_1} 
					+ \frac{\bar\sigma^2_2}{\sigma^2_2} \right)\right] \right\},
\end{equation}
where $\bar\sigma_1^2$ and $\bar\sigma_2^2$ are defined by 
\begin{equation} \label{sigma bar def}
\bar\sigma^2_i = \frac{1}{N} \sum_{k=1}^N \left(h_i^k\right)^2 
\end{equation} 
for $i=1,2$.
It is easily shown that the maximum in Eq.~(\ref{denominator to maximize}) is achieved at
$\sigma_i = \bar\sigma_i$.
From Eq.~(\ref{general likelihood estimator}) this yields
\begin{widetext}
\begin{equation} 
\Lambda_\text{ML} = \frac{\displaystyle \max_{{\bf v}_s\in\Theta_{s1}}~\max_{{\bf v}_n\in\Theta_n}~ \int d^{ND}s ~p_{\mathcal{N|V}_n}(h-s|{\bf v}_n) p_{\mathcal{S|V}_s,\mathcal{T}}(s|{\bf v}_s,1) }
                         {\displaystyle \left( 2\pi \bar\sigma_1 \bar\sigma_2 \right)^{-N} \exp\left( -N\right) }.
\end{equation}
Combining this with Eq.~(\ref{assumption4}) yields the following final general expression for the maximum likelihood statistic:
\begin{equation} \label{special likelihood estimator}
\Lambda_\text{ML} = \max_{{\bf v}_s\in\Theta_{s1}}~\max_{\sigma_1 \ge 0}~\max_{\sigma_2 \ge 0}~
		    \prod_{k=1}^N  \frac{ \bar\sigma_1 \bar\sigma_2 }{ \sigma_1 \sigma_2 }   
		    \int_{-\infty}^\infty ds^k~ p_{\mathcal{S}^k|\mathcal{V}_s,\mathcal{T}}(s^k|{\bf v}_s,1) 
	            \exp\left[ -\frac{ \left(h_1^k - s^k \right)^2 }{ 2\sigma^2_1 } 
	                       - \frac{ \left( h_2^k - s^k \right)^2 }{ 2\sigma^2_2 } 
	       	               + 1 \right].
\end{equation} 
\end{widetext}

\subsection{Gaussian signal}
\label{ss:Gaussian signal}

We now consider the case where the signal is Gaussian and has a
vanishing mean.  We denote by $\alpha^2$ the variance of the signal,
so the prior for $\mathcal{S}$ is given by
\begin{equation} \label{Gaussian signal}
p_{\mathcal{S}^k|\mathcal{V}_s,\mathcal{T}}(s^k|\alpha,1) = \frac{ 1 }{ \sqrt{2 \pi} \alpha } 
			      \exp\left[- \frac{ \left( s^k \right)^2 }{ 2 \alpha^2 } \right].
\end{equation}
For this model ${\bf v}_s = (\alpha)$ has only one component, and
$\Theta_{s1}=\{ \alpha ~|~ \alpha > 0\}$.

Substituting the signal probability distribution (\ref{Gaussian signal}) into the general expression (\ref{special likelihood estimator}) 
for $\Lambda_\text{ML}$ yields a Gaussian integral which is straightforward to evaluate.  The result is
\begin{eqnarray} 
\label{long Gaussian stat}
\Lambda^\text{G}_\text{ML} &=&  \max_{\alpha > 0}~\max_{\sigma_1 \ge 0}~\max_{\sigma_2 \ge 0} 
\left\{ 
 \frac{\bar\sigma_1 \bar\sigma_2}{\sqrt{\sigma_1^2 \sigma_2^2 + \sigma_1^2 \alpha^2 + \sigma_2^2 \alpha^2}}   \right. \\
&\times & \left. \exp\left[ \frac{ \frac{\bar\sigma_1^2}{\sigma_1^4} + \frac{\bar\sigma_2^2}{\sigma_2^4} + \frac{2\bar\alpha^2}{\sigma_1^2\sigma_2^2} }
	                   {2 \left( \frac{1}{\sigma_1^2} + \frac{1}{\sigma_2^2} + \frac{1}{\alpha^2} \right)} 
		           - \frac{\bar\sigma_1^2}{2\sigma_1^2} - \frac{\bar\sigma_2^2}{2\sigma_2^2} + 1 \right] \right\}^N ,\nonumber 
\end{eqnarray} 
where 
\begin{equation} 
\bar \alpha^2 = \frac{1}{N}\sum_{k=1}^N h_1^k h_2^k,
\label{baralphadef}
\end{equation} 
and we have appended a superscript G on $\Lambda^\text{G}_\text{ML}$ to indicate the maximum likelihood detection 
statistic for a Gaussian signal.  

One can show that the maximum in Eq.~(\ref{long Gaussian stat}) is achieved at $\alpha = \hat\alpha$, $\sigma_1  = \hat\sigma_1$, and
$\sigma_2  = \hat\sigma_2$, where
\begin{eqnarray}
\hat\alpha^2 &=& \bar\alpha^2 ~\theta(\bar\alpha^2) ,  \label{gaussian estimator1}\\
\hat\sigma^2_i &=& (\bar\sigma^2_i - \hat\alpha^2) ~\theta\left( \bar\sigma^2_i - \hat\alpha^2 \right) , \label{gaussian estimator2}
\end{eqnarray} 
for $i=1,2$, and $\bar\sigma_1$ and $\bar\sigma_2$ are given by Eq.~(\ref{sigma bar def}). 
Here $\theta(x)$ is the step function (\ref{stepfunction}). 
The quantities (\ref{gaussian estimator1}) and (\ref{gaussian estimator2}) are the maximum likelihood estimators for 
the variance $\alpha^2$ of the signal and the variances $\sigma_1^2$
and $\sigma_2^2$ of the noise in the two detectors. The step functions  
in Eqs.~(\ref{gaussian estimator1}) and (\ref{gaussian estimator2})
arise as a result of the bounds of the maximization 
in Eq.~(\ref{special likelihood estimator}).

The corresponding detection statistic is, from Eq.~(\ref{long Gaussian stat})
% FOOTNOTE %%%%%%%%%%%%%%%%%%%%%%%%%%%%%%%%%%%%%%%%%%%%%%%%%%%%%%%%%%%%%%%%%%%%%%%%%%%%%%%%%%%%%%%%%%%%%%%%%%
\footnote{To simplify the formula for $\Lambda_\text{ML}^\text{G}$ we assume that $\bar\sigma_i^2 - \bar\alpha^2> 0$.  
This will be true for any realistic value of $N$ since $\bar\sigma_i^2
- \bar\alpha^2 = \sigma_{i,{\rm true}}^2 + O(1/\sqrt{N})$, where
$\sigma_{i,{\rm true}}$ is the true value of $\sigma_i$ and the second term
describes the statistical fluctuations.},
%%%%%%%%%%%%%%%%%%%%%%%%%%%%%%%%%%%%%%%%%%%%%%%%%%%%%%%%%%%%%%%%%%%%%%%%%%%%%%%%%%%%%%%%%%%%%%%%%%%%%%%%%%%%%
\begin{equation} \label{Gaussian statistic}
\Lambda^\text{G}_\text{ML} = \left[ 1 - \frac{\hat\alpha^4}{\bar\sigma_1^2 \bar\sigma_2^2} \right]^{-N/2}.
\end{equation} 
The cross-correlation statistic $\Lambda_\text{CC}$ can be obtained
from $\Lambda_\text{ML}^\text{G}$ via a monotonic transformation which
preserves false dismissal versus false alarm curves [cf.\ Eq.\
(\ref{transformation}) above]:
\begin{eqnarray}
\Lambda_\text{CC} &=& \sqrt{1 -(\Lambda^\text{G}_\text{ML})^{-2/N}}
\nonumber \\
 &=& \frac{\hat\alpha^2}{\bar\sigma_1 \bar\sigma_2}.
\label{standard cross corr}
\end{eqnarray}

Note that if we had assumed the noise parameters ${\bf v}_n =
(\sigma_1,\sigma_2)$ were known, and derived a statistic from
Eq.~(\ref{known likelihood estimator}) rather than Eq.~(\ref{general
likelihood estimator}), we would have found instead the detection
statistic $\tilde\Lambda_\text{ML}^\text{G} =
\bar\Lambda_\text{ML}^\text{G}~\theta(\bar\Lambda_\text{ML}^\text{G})$,
where 
\begin{equation} \label{known varriance}
\bar\Lambda_\text{ML}^\text{G} = \bar\alpha^2 + \frac{1}{2}\left[
                    \frac{\sigma_2^2}{\sigma_1^2}(\bar\sigma_1^2 -
                    \sigma_1^2)                     +
                    \frac{\sigma_1^2}{\sigma_2^2}(\bar\sigma_2^2 -
                    \sigma_2^2)  \right], 
\end{equation}
which is different from the standard cross-correlation statistic. This
non-standard result is obtained because of the unrealistic assumption 
that the noise parameters ${\bf v}_n = (\sigma_1,\sigma_2)$ are
known.  Different derivations of the result (\ref{known varriance})
can be found in Refs.~\cite{sam joe,robust gaussian II}.

It is often useful to characterize the ``strength'' of a stochastic
background in terms of the signal-to-noise ratio of the
cross-correlation statistic (\ref{standard cross corr}), which we now
define.  First note that for large $N$, the fractional fluctuations in 
$\hat\alpha^2$ will be much larger than those in
$\bar\sigma_1\bar\sigma_2$
\footnote{This is true at fixed signal-to-noise ratio $\rho$.}.  
For the purpose of defining the signal-to-noise ratio, we assume that $N$
is large enough that  
$\bar\sigma_1$ and $\bar\sigma_2$ in Eq.~(\ref{standard cross corr})
can be taken to be independent of $h$, 
so that $\Lambda_\text{CC}$ and $\hat\alpha^2$ are equivalent
detection statistics.   
We also use ${\bar \alpha}^2$ instead of
${\hat \alpha}^2$ in the computations that follow, as is conventional
when defining 
signal-to-noise ratios.  If a signal is present, then the expected
value of $\bar\alpha^2$ is, from Eqs.\ (\ref{detector output
matrices}), (\ref{assumption2})--(\ref{assumption4}), (\ref{Gaussian
signal}) and (\ref{baralphadef}),
\begin{equation} \label{expected value 1}
\left< \bar \alpha^2 \right> = \alpha^2.
\end{equation} 
If no signal is present, so that $\alpha^2=0$, then the fluctuations in $\bar\alpha^2$ are given by
\begin{equation} \label{fluctuations 1}
\Delta \left( \bar \alpha^2 \right) = \frac{\sigma_1\sigma_2}{\sqrt{N}}.
\end{equation} 
The signal-to-noise ratio $\rho$ is defined to be the ratio of these two quantities:
\begin{equation} \label{rho}
\rho = \frac{\alpha^2\sqrt{N}}{\sigma_1\sigma_2}.
\end{equation} 

\subsection{Non-Gaussian signal}
\label{ss:Non-Gaussian signal}

As mentioned in the introduction, the traditional assumption that a
gravitational wave stochastic background will be Gaussian  
requires the individual events to be sufficiently frequent and
uncorrelated.  Our model for a non-Gaussian signal assumes instead that the
events are infrequent.  

\begin{figure}
\begin{center}
\epsfig{file=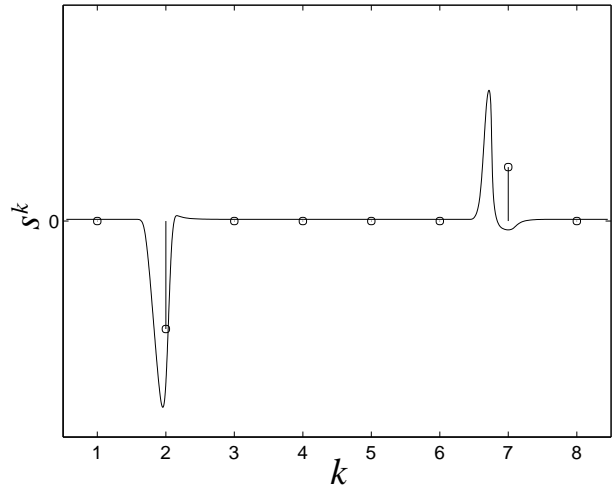,width=8cm}
\caption{
Sketched segment of the signal produced by a model non-Gaussian stochastic
background of events unresolved by the detectors. Here we show two events.  The solid curve is the 
exact signal.  This exact signal's contributions to the detector outputs, shown as stemmed {\sf o}'s, 
are averages of the exact signal over the detector resolution timescale.
}
\label{signal sketch}
\end{center}
\end{figure}

Consider a collection of similar events generating a stochastic background $\mathcal{S}$. 
Let $\xi$ be the probability that, at any randomly chosen time, the waves from an event 
are arriving at the detectors.  We assume that
the time structure of individual
events cannot be resolved by the detectors.  That is, 
we assume that the events occur over timescales smaller than the
detectors' resolution time, as illustrated in Fig.~\ref{signal sketch}.
We assume that the distribution of the amplitudes of the events is
Gaussian with variance $\alpha^2$.
The probability distribution for the signal given the signal
parameters $(\xi,\alpha)$ is therefore given by
\begin{eqnarray} \label{signal model}
p_{\mathcal{S}^k|\mathcal{V}_s,\mathcal{T}}[s^k| (\xi,\alpha),1] &=& 
\frac{\xi}{\sqrt{2\pi}\alpha}\exp \left[ -\frac{\left( s^k\right) ^2}{2\alpha^2} \right] \nonumber \\
&+& (1-\xi) \delta \left( s^k \right) \label{signal prior},
\end{eqnarray} 
together with Eq.~(\ref{assumption4}).
Thus the signal model parameters are ${\bf v}_s=(\xi,\alpha)$, which
give respectively the  ``event probability'' and ``event variance''
characterizing the stochastic background.  The parameter space  
$\Theta_s$ for this model is 
\begin{equation} 
\Theta_{s} = \left\{ (\xi,\alpha) ~|~ 0 \le \xi \le 1 \text{ and } \alpha \ge 0 \right\},
\end{equation} 
and the subset corresponding to a signal being present is
\begin{equation} 
\Theta_{s1} = \left\{ (\xi,\alpha) ~|~ 0 < \xi \le 1 \text{ and } \alpha > 0 \right\}.
\end{equation}

Note that our assumption that the time structure of events is not resolved by the detector is unrealistic.  Detector resolution times 
can be as small as 0.1 ms in the case of ground-based detectors like LIGO
\footnote{For ground-based detectors, the effective resolution time in a cross-correlation between
two detectors can be considerably longer than $0.1$ ms \cite{Allen
  Romano}, which may help with this issue.},
and even supernova bursts are expected to
have time scales $\gtrsim 10$ ms \cite{waveform catalog,new waveform catalog}.  
It will be important for future studies to relax this assumption.

We now compute the maximum likelihood detection statistic $\Lambda^\text{NG}_\text{ML}$ for our simple non-Gaussian signal model
by substituting Eq.~(\ref{signal prior}) into Eq.~(\ref{special likelihood estimator}).
This yields
\begin{widetext}  
\begin{eqnarray}  \label{main result2}
\Lambda_{\text{ML}}^{\text{NG}} &=&
\max_{0<\xi\le 1}~ \max_{\alpha > 0}~ \max_{\sigma_1 \ge 0}~ \max_{\sigma_2 \ge 0}~ \prod_{k=1}^N
\left\{  
        \frac{ \bar\sigma_1 \bar\sigma_2 \xi}{\sqrt{\sigma^2_1 \sigma^2_2 + \sigma^2_1 \alpha^2 + \sigma^2_2 \alpha^2}} 
        \exp \left[ \frac{\left( \frac{h_1^k}{\sigma^2_1} + \frac{h_2^k}{\sigma^2_2}\right)^2}
        {2\left( \frac{1}{\sigma^2_1} + \frac{1}{\sigma^2_2} + \frac{1}{\alpha^2} \right)} 
        - \frac{\left( h_1^k\right)^2}{2\sigma^2_1} - \frac{\left( h_2^k\right)^2}{2\sigma^2_2} + 1\right]  \right. \nonumber \\ 
&+& \left. \frac{\bar\sigma_1 \bar\sigma_2}{\sigma_1 \sigma_2}  (1-\xi)
        \exp \left[ - \frac{\left( h_1^k\right)^2}{2\sigma^2_1} - \frac{\left( h_2^k\right)^2}{2\sigma^2_2} + 1\right]\right\}.
\end{eqnarray}
\end{widetext}
The values of $\xi$, $\alpha^2$, $\sigma_1^2$, and $\sigma_2^2$ which achieve the maximum in Eq.~(\ref{main result2})
are, respectively, estimators of the signal's Gaussianity parameter,
the variance of the signal events, 
and the noise variances in the two detectors\footnote{See Ref.~\cite{MG9} for a derivation of a statistic similar to $\Lambda_\text{ML}^\text{NG}$ and
designed for the same non-Gaussian signals which is based on Eq.~(\ref{known likelihood estimator}) rather
than Eq.~(\ref{general likelihood estimator}).}.
%%%%%%%%%%%%%%%%%%%%%%%%%%%%%%%%%%%%%%%%%%%%%%%%%%%%%%%%%%%%%%%%%%%%%%%%%%%%%%%%%%%%%%%%%%%%%%%%%%%%%%%%%%%%%
Note that if we evaluate Eq.~(\ref{main result2}) at $\xi=1$, rather than maximizing over $\xi$, 
we recover Eq.~(\ref{long Gaussian stat}) and the statistic $\Lambda_\text{ML}^\text{G}$.

We mention in passing an approximate version of the statistic
(\ref{main result2}) which is significantly easier to compute.
Expanding the logarithm of the 
quantity to be maximized in Eq.~(\ref{main result2}) as a power 
series in $\alpha^2$ to fourth order about $\alpha^2=0$ yields the 
approximate statistic $\hat\Lambda_\text{ML}^\text{NG}$ given by 
\begin{eqnarray} \label{expanded}
\ln \hat\Lambda_\text{ML}^\text{NG} &=& \max_{0<\xi\le 1}~ \max_{\alpha > 0}~ \max_{\sigma_1 \ge 0}~ \max_{\sigma_2 \ge 0}~
                        \sum_{n=0}^4 
                        \sum_{l = 0}^{8} 
                        \sum_{m=0}^8 
                        \left(\frac{\alpha^2}{\sigma_1 \sigma_2}\right)^n \nonumber \\
&\times&		C_{nlm}\left(\xi,\sigma_1,\sigma_2\right) 
			\sum_{k=1}^N (h_1^k)^l (h_2^k)^m,
\end{eqnarray} 
where the coefficients $C_{nlm}(\xi,\sigma_1^2,\sigma_2^2)$ 
%are tabulated in Appendix \ref{s:coeffs}.  These coefficients 
vanish
unless $l+m$ is even and $l+m \le 8$. 
In evaluating the statistic (\ref{expanded}), one can first evaluate the 24 sums
\begin{equation} 
\sum_{k=1}^N (h_1^k)^l (h_2^k)^m
\end{equation} 
for the required values of $l$ and $m$, and subsequently numerically maximize over the parameters $\xi$, 
$\alpha$, $\sigma_1$, and $\sigma_2$. Thus the length-$N$ sums need only be performed once, rather than each time one tries a new 
set of values for $\xi$, $\alpha$, 
$\sigma_1$, and $\sigma_2$. Therefore the computational cost of $\hat\Lambda_\text{ML}^\text{NG}$ is only about an order of 
magnitude greater than that of the cross correlation statistic
$\Lambda_\text{CC}$, and this statistic may be useful to explore.

We now derive the signal-to-noise ratio $\rho$ for the cross-correlation
statistic and for the non-Gaussian signal (\ref{signal model}).
If the signal is present, then from Eqs.\ (\ref{detector output
matrices}), (\ref{assumption4}),
(\ref{baralphadef}), (\ref{gaussian estimator1}) and (\ref{signal model})
the expected value
of $\bar\alpha^2$ is
\begin{equation} \label{expected value 2}
\left< \bar \alpha^2 \right> = \xi\alpha^2.
\end{equation}
If no signal is present then the fluctuations in $\bar\alpha^2$ are given by
\begin{equation} \label{fluctuations 2}
\Delta \left( \bar \alpha^2 \right) = \frac{\sigma_1\sigma_2}{\sqrt{N}}.
\end{equation}
Therefore, taking the ratio of Eqs.\ (\ref{expected value 2}) and
(\ref{fluctuations 2}), the signal-to-noise ratio $\rho$ is
\begin{equation} \label{rho2}
\rho = \frac{\xi\alpha^2\sqrt{N}}{\sigma_1\sigma_2}.
\end{equation}

\section{Performance comparison}
\label{s:Performance comparison}

In this section we compare the performances of the cross-correlation
statistic (\ref{standard cross corr}), the burst statistic
(\ref{eq:lambdaBdef}), and the maximum
likelihood statistic (\ref{main result2})
for our model non-Gaussian signal described in Sec.~\ref{ss:Non-Gaussian signal}.   
The comparison is quantified in terms of the false alarm versus false
dismissal curves, as discussed in Sec.\ \ref{s:General theory of detection
statistics and parameters estimator} above.
In Sec.\ \ref{ss:analytic} we discuss analytic predictions for these curves
for the three different statistics.  Section \ref{ss:Description of the
simulation algorithm} describes our Monte Carlo simulation algorithm,
and Secs.\ \ref{ss:Results for detection} and \ref{ss:Results for
  parameter estimation} describe the results.

\subsection{Analytic computation of asymptotic behavior of statistics}
\label{ss:analytic}

We start by discussing the set of parameters on which the false dismissal versus false 
alarm curves can depend.  As before, we assume two detectors with noise characterized by
Eq.~(\ref{assumption2}) with $\mathcal{V}_n=(\sigma_1,\sigma_2)$,  and a non-Gaussian 
signal characterized by Eqs.~(\ref{assumption4}) and (\ref{signal prior}) with 
$\mathcal{V}_s=(\xi,\alpha)$.
The curves for each statistic are given by some function
\begin{equation}\label{dependance1}
P_\text{FD} = P_\text{FD}(P_\text{FA},\xi,\alpha,\sigma_1,\sigma_2,N)
\end{equation}
of the false alarm probability $P_{\rm FA}$, the Gaussianity parameter
$\xi$, the rms amplitude $\alpha$ of events, the noise variances
$\sigma_1^2$ and $\sigma_2^2$, and the number of data points $N$.
We can simplify Eq.~(\ref{dependance1}) by replacing $\alpha$ with the
signal-to-noise ratio $\rho$ using the definition (\ref{rho2}), and
noting from dimensional analysis that $P_\text{FA}$ depends on $\sigma_1$ 
and $\sigma_2$ at fixed $\rho$ only through the ratio
$\sigma_1/\sigma_2$.  This gives  
\begin{equation}\label{dependance2}
P_\text{FD} = P_\text{FD}(P_\text{FA},\xi,\rho,\sigma_1/\sigma_2,N).
\end{equation}
For simplicity, we specialize to $\sigma_1=\sigma_2$ for the remainder of this paper.  
This implies that
\begin{equation}\label{dependance3}
P_\text{FD} = P_\text{FD}(P_\text{FA},\xi,\rho,N).
\end{equation}

\subsubsection{Cross correlation statistic}

The false dismissal versus false alarm curves for the cross-correlation statistic can be computed 
analytically in the large $N$ limit, as we now describe.  Our derivation generalizes the analysis of 
Ref.~\cite{Allen Romano} from Gaussian to non-Gaussian signals. For any detection statistic $\Gamma$, 
we can express $P_\text{FA}$ and $P_\text{FD}$ in terms of the detection threshold $\Gamma_*$ as
\begin{eqnarray} 
P_\text{FA}(\Gamma_*,\sigma_1,\sigma_2,N) &=& 
         \int_{\Gamma_*}^\infty dx~p_{\Gamma|\mathcal{T}}(x|0) , \label{simple Pfa}\\
P_\text{FD}(\Gamma_*,\xi,\rho,\sigma_1,\sigma_2,N) &=& 
         1 - \int_{\Gamma_*}^\infty dx~p_{\Gamma|\mathcal{T}}(x|1)\nonumber  . \label{simple Pfd}\\
\end{eqnarray}
Here the definition of the random variable $\mathcal{T}$ is such that
if $\mathcal{T}=0$ then no signal is present ($\xi = \rho = 0$), and  
if $\mathcal{T}=1$ then a signal is present ($\xi \ne 0$ and $\rho \ne
0$); cf.\ Sec.\ \ref{ss:Notational conventions} above.
Note that by eliminating $\Gamma_*$ between
Eqs.~(\ref{simple Pfa}) and (\ref{simple Pfd}), we recover Eq.~(\ref{dependance1}).

In the large $N$ limit, the distribution
$p_{\Lambda_\text{CC}|\mathcal{T}}(x|t)$ is a Gaussian by the
central limit theorem, and the integrals
(\ref{simple Pfa}) and (\ref{simple Pfd}) can be evaluated
analytically (see Appendix \ref{s:appendixB}) to give
\begin{eqnarray} \label{analytic}
&&P_\text{FD} \left(P_\text{FA},\xi,\rho,N \right) =  1  \\
&&-\frac{1}{2} \erfc \left[ \frac{\displaystyle \erfc^{-1} \left(2P_\text{FA}\right) - \frac{\rho}{\sqrt{2}} }  
                           {\sqrt{\displaystyle \frac{\rho^2}{N}\left( \frac{3}{\xi} - 1 \right) + \frac{2\rho}{\sqrt{N}} + 1}} 
		      \right] + O \left( {1 \over \sqrt{N}} \right). \nonumber
\end{eqnarray} 
Here the function $\erfc(x)$ (known as the compliment of the error
function) is defined by 
\begin{equation} 
\erfc(x) = \frac{2}{\sqrt{\pi}}\int_x^\infty dy~e^{-y^2},
\end{equation}
and $\erfc^{-1}(x)$ is the inverse of $\erfc(x)$.  
The formula (\ref{analytic}) is valid only for $P_{\rm FA} < 1/2$; 
$P_\text{FD}$ is undefined for $1/2 \le P_\text{FA} < 1$.  
In deriving Eq.~(\ref{analytic}), we assumed 
that the statistics $\Lambda_\text{CC}$ and $\hat\alpha^2$ 
are equivalent, and that the distribution for $\bar\alpha^2$ is
Gaussian. Those assumptions  
are only valid up to fractional correction terms of order
$1/\sqrt{N}$; hence the indicated correction term in Eq.\ (\ref{analytic}).

\begin{figure}
\begin{center}
\epsfig{file=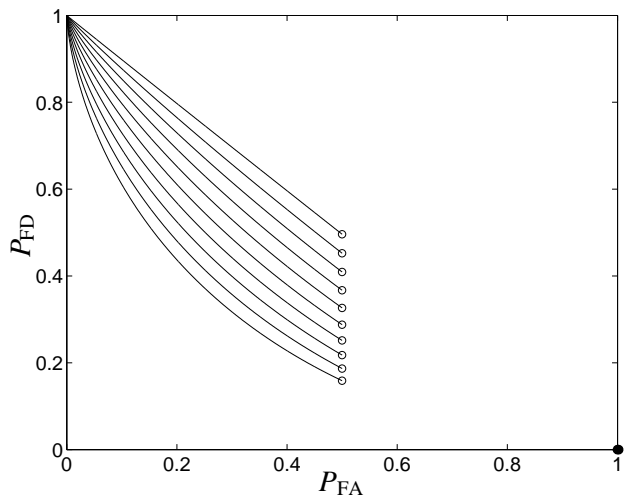,width=8.5cm}
\caption{Sample false dismissal versus false alarm curves for
the cross correlation statistic $\Lambda_\text{CC}$ 
in the large $N$ limit, as prescribed by Eq.~(\ref{specialized
  analytic}).  For these curves  
the signal-to-noise ratio $\rho$ has equally spaced values from 0.01
to 1. Note that here $P_\text{FD}$  
is undefined for $1/2 \le P_\text{FA} <  1$.}
\label{analytical curves}
\end{center}
\end{figure}

In the regime where $\rho^2\ll N \xi$ in addition to $N \gg 1$, the
result (\ref{analytic}) simplifies to 
\begin{eqnarray} 
P_\text{FD} \left(P_\text{FA},\xi,\rho,N \right) &=&  1 - \frac{1}{2} 
\erfc \left[ \erfc^{-1} \left(2P_\text{FA}\right) -
\frac{\rho}{\sqrt{2}} \right]  \nonumber \\ 
&+& O\left( \frac{1}{ \sqrt{N} }\right) + O\left({ \rho \over
\sqrt{N}} \right) + O\left({ \rho^2 \over
N \xi } \right). \nonumber \\
 \label{specialized analytic}
\end{eqnarray}
Note that the false dismissal versus false alarm relation 
(\ref{specialized analytic}) is independent of both $N$ and $\xi$.
Sample curves from Eq.~(\ref{specialized analytic}) are shown in
Fig.~\ref{analytical curves}.   
The discontinuities at $P_\text{FA} = 1/2$ are a result of the step
functions in the definition (\ref{gaussian estimator1})
of $\hat\alpha^2$.

\subsubsection{Burst statistic}

By combining the definition (\ref{eq:lambdaBdef}) of the burst
statistic together with the decomposition (\ref{eq:common}), the
noise and signal distributions (\ref{assumption2}) and (\ref{signal
prior}), and the change of variables (\ref{rho2})
it is straightforward to derive the exact false alarm versus
false dismissal relation.  The result is given by
\begin{eqnarray}
(1 - P_{\rm FA})^{1/N} = {\rm erf}\left({\Lambda_* \over \sqrt{2}}\right)
\label{burstans1}
\end{eqnarray}
and
\begin{equation} 
P_{\rm FD}^{1/N} = \xi \, {\rm erf} \left[ { \Lambda_* \over \sqrt{
       2 + {2 \rho \over \xi \sqrt{N}} }} \right] + (1 -
      \xi) \, {\rm erf} \left( {\Lambda_* \over \sqrt{2}} \right),
\label{burstans2}
\end{equation}
where $\Lambda_*$ is the value of the threshold.

\subsubsection{Maximum likelihood statistic}
\label{s:MLS}

We start by discussing the
different regimes present in the space of signal 
parameters $\xi$, $\rho$ and $N$, treating the 
false alarm probability $P_\text{FA}$ as fixed.  There are several
different constraints  
on the three parameters $\xi$,
$\rho$, and $N$ that define the regime in parameter space where we
expect our maximum likelihood statistic to work well. First, it is clear that the total 
number of events $\sim \xi N$ in the data set must be large compared to one:
\begin{equation}
\xi \gg \frac{1}{N}.
\label{eq:constraint2a}
\end{equation}

Second, if the signal-to-noise ratio $\alpha^2 / (\sigma_1 \sigma_2)$
of individual burst events is large compared to one, then one can detect the
individual events using the burst statistic (\ref{eq:lambdaBdef})
and the method of this paper is not needed.  From
Eq.\ (\ref{rho}) we can write the constraint $\alpha^2 / (\sigma_1
\sigma_2) \alt 1$ as
\begin{equation}
\xi \agt \frac{\rho}{\sqrt{N}}.
\end{equation}
A more precise version of this requirement can be obtained by noting
that the detection threshold for the signal-to-noise ratio 
$\alpha^2 / (\sigma_1 \sigma_2)$ is $\sim \sqrt{2 \ln N}$, since there
are $N$ independent trials.  This yields the constraint
\begin{equation}
\xi \agt \frac{\rho}{\sqrt{2 N \ln N}}.
\label{eq:constraint2}
\end{equation}
The regime $\xi \sim \rho / \sqrt{2 N \ln N}$ is where the 
burst statistic $\Lambda_\text{B}$ starts becoming as sensitive as the
cross correlation statistic, as can be seen by combining Eqs.\
(\ref{specialized analytic}), (\ref{burstans1}) and (\ref{burstans2})
above.  This behavior can also be seen  
in Figs. \ref{omega gain} and \ref{fig:theoretical} above.

A third constraint on the space of signal parameters is derived as
follows.  Consider the statistic 
\begin{equation}
\eta = \frac{1}{N}\sum_{k=1}^N (h_1^k)^2(h_2^k)^2.
\label{etadef}
\end{equation} 
We can use this statistic to estimate the Gaussianity parameter $\xi$
in the following way.
The mean value of $\eta$ when a signal is present is given by
\begin{equation} \label{eta mean}
\left< \eta \right> = 3\xi\alpha^4 + \xi\alpha^2(\sigma_1^2+\sigma_2^2) + \sigma_1^2 \sigma_2^2,
\end{equation} 
and the variance when a signal is absent is 
\begin{equation} \label{eta var}
(\Delta \eta)^2 = \sigma^4_1\sigma^4_2 \frac{8}{N}.
\end{equation} 
It follows from Eqs.\ (\ref{eta mean}), (\ref{eta var}), and the
relation $\left<
\hat\alpha^2 \right> = \xi\alpha^2$ that the estimator ${\hat \xi}$ of
$\xi$ defined by
\begin{equation}
\hat \xi = \frac{3 \hat\alpha^4}{\eta - \hat\alpha^2 (\hat\sigma_1^2 + \hat\sigma_2^2) - \hat\sigma_1^2\hat\sigma_2^2} 
\end{equation} 
has a fractional accuracy of order
\begin{equation}
\frac{\Delta \xi}{\xi} \sim \frac{\xi \sqrt{N}}{\rho^2}.
\label{Deltaxi}
\end{equation}
Now in the regime $\Delta \xi / \xi \ll 1$, we expect our maximum
likelihood detection statistic to work well, since one's first guess
for a nonlinear statistic (\ref{etadef}) can be used to detect the
non-Gaussianity of the signal to high accuracy.
In the regime $\Delta \xi / \xi \gg 1$, it is not obvious how the
maximum likelihood detection statistic will perform, since it could
have a performance much better than that of the statistic $\eta$.
However, our Monte Carlo simulations [Sec.\ \ref{ss:Description of the
simulation algorithm} below] and analytic 
computations [Appendix 
\ref{s:appendixC}] indicate that the maximum likelihood statistic does
indeed perform poorly in the regime $\Delta \xi / \xi \gg 1$.
Thus, our third constraint is $\Delta \xi / \xi \alt 1$, which from
Eq.\ (\ref{Deltaxi}) can be written as 
\begin{equation}
\xi \alt \frac{\rho^2}{\sqrt{N}}.
\label{constraint3}
\end{equation}
Our Monte Carlo simulations show that for $\rho^2/\sqrt{N} \alt \xi
\alt 1$, the maximum likelihood and cross-correlation statistics
perform roughly equivalently, and that once $\xi$ becomes smaller than
$\rho^2 / \sqrt{N}$, the maximum likelihood statistic starts to
perform significantly better than the cross-correlation statistic; see 
Figs. \ref{omega gain} and \ref{fig:theoretical} above.

In Appendix \ref{s:appendixC} we derive analytically the approximate
expression (\ref{eq:ansA}) for the false dismissal
probability for the maximum likelihood statistic, which we expect to
be accurate up to corrections of order $1/\rho^4$ or a few tens of
percent.  We also derive the expression (\ref{eq:ansB}) for the false
alarm probability using a combination of analytical and numerical
techniques.  Combining these results gives the curves which are associated with the
maximum likelihood statistic $\Lambda_\text{ML}^\text{NG}$ and labeled ``analytic'' in 
Figs.~\ref{omega gain}, \ref{fig:theoretical}, \ref{detection curves}, and \ref{LargeN}.

\subsection{Description of the Monte Carlo simulation algorithm}
\label{ss:Description of the simulation algorithm}

Next we describe our Monte Carlo simulations of the performances of the various statistics.
We numerically estimate the false dismissal and false alarm probabilities
$P_\text{FD}$ and $P_\text{FA}$ by conducting an ensemble
of $N_E$ simulated experiments.  
For each experiment we simulate a detector output matrix, half of which 
have a signal present, and half of which do not.  Since we know in advance whether or not a signal is present, we can 
easily estimate $P_\text{FA}$ and $P_\text{FD}$.  More specifically,
our algorithm for simulating false  
dismissal versus false alarm curves, for an arbitrary statistic
$\Gamma$, is as follows:
\begin{enumerate}
\item Choose values for $\xi$, $\alpha$, $\sigma_1$, $\sigma_2$, and $N$.
\item Choose the total number of trials $N_E$.
\item For $r=1,2,\ldots,N_E/2$:
	\begin{enumerate}
                \item Generate a data train $h(\sigma_1,\sigma_2,N)$ of noise only.
                \item Compute $\Gamma$ and store result as $\Gamma_{r0}$.
		\item Generate a data train $h(\xi,\alpha,\sigma_1,\sigma_2,N)$ which has a signal present.
		\item Compute $\Gamma$ and store result as $\Gamma_{r1}$.
	\end{enumerate}
\item Choose a discretization $\Gamma_{*j}$ of the set of thresholds,
where $j=1,2,\ldots,M$.
\item Set $P_\text{FA}(\Gamma_{*j}) = P_\text{FD}(\Gamma_{*j}) = 0$, for each $j$.
\item For $r=1,2,\ldots,N_E/2$:
	\begin{enumerate}
		\item for each $j$, if $\Gamma_{r0} > \Gamma_{*j}$, increment $P_\text{FA}(\Gamma_{*j})$ by $2/N_E$.
		\item for each $j$, if $\Gamma_{r1} \le \Gamma_{*j}$, increment $P_\text{FD}(\Gamma_{*j})$ by $2/N_E$.
	\end{enumerate}
\item Repeat steps 3-6 above several times to estimate the fluctuations in $P_\text{FA}(\Gamma_{*j})$ and $P_\text{FD}(\Gamma_{*j})$.
\end{enumerate}

We use the above algorithm to simulate false dismissal versus false
alarm curves for the three statistics $\Lambda_\text{CC}$,
$\Lambda_{\rm B}$ and 
$\Lambda_\text{ML}^\text{NG}$.  The analytical expressions
(\ref{analytic}) and (\ref{burstans1}) -- (\ref{burstans2}) for the
cross-correlation and burst statistics are 
used as a check of the numerical method.

\subsection{Simulation results}
\label{ss:Results for detection}

\begin{figure}
\begin{center}
\epsfig{file=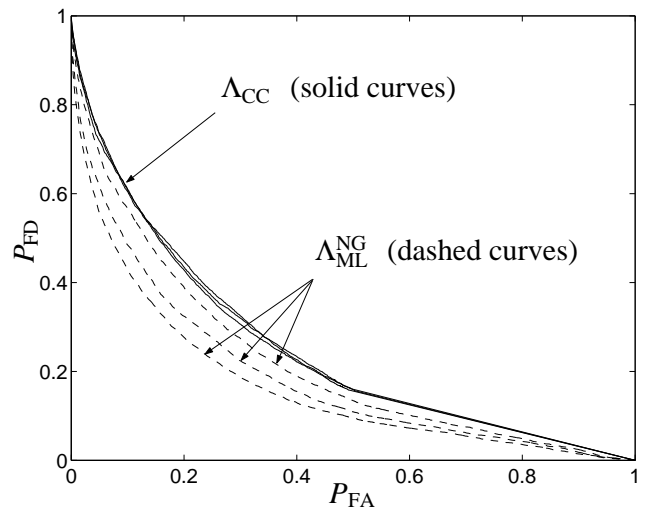,width=8.5cm}
\caption{
Plots of false dismissal probability ($P_\text{FD}$) versus false alarm
probability ($P_\text{FA}$) for the standard cross-correlation
statistic $\Lambda_\text{CC}$ and 
our maximum likelihood statistic $\Lambda_\text{ML}^\text{NG}$.  Each
of these curves is characterized by a total number of trials $N_E =
2\times 10^4$, number of data points $N = 5\times10^4$, noise variances
$\sigma_1 = \sigma_2 = 1$, and by the signal-to-noise ratio $\rho = 1$.
The values of the Gaussianity parameter $\xi$ are 0.02, 0.012, and
0.01.  The solid curves are the results for $\Lambda_{\rm ML}^{\rm
G}$; these curves are bunched together because  
$\rho$ is fixed.  The dashed curves are the results for $\Lambda_{\rm
ML}^{\rm NG}$.  For the dashed curves, the lowest curve is for $\xi =
0.01$, while the highest curve is for $\xi = 0.02$. 
We estimate error bars for each of these curves by separating the $2 \times
10^4$ runs into 10 bins of $2 \times 10^3$, and generating 10 separate
plots; the resulting fluctuations are $\alt 10^{-3}$.  The curves for
the cross correlation statistic $\Lambda_{\rm ML}^{\rm G}$ agree with
the analytic prediction (\ref{analytic}) to within $\sim 10^{-3}$.
This plot shows that $\Lambda_\text{ML}^\text{NG}$ can perform 
significantly better than $\Lambda_\text{CC}$.
}
\label{compare}
\end{center}
\end{figure}

A family of simulated false dismissal versus false alarm curves for
the cross correlation statistic $\Lambda_\text{CC}$ and
the maximum likelihood statistic $\Lambda_\text{ML}^\text{NG}$ is  
shown in Fig.~\ref{compare}. 
We see that at fixed $\rho$, as the Gaussianity $\xi$ of the signal
decreases, $\Lambda_\text{ML}^\text{NG}$ performs increasingly better
than $\Lambda_\text{CC}$.  
The curves for $\Lambda_\text{CC}$ are almost indistinguishable from
each other because $\rho$ is fixed, and the curves depend only on
$\rho$ and not on $\xi$ for this detection statistic in the large $N$
limit [cf.\ Eq.\ (\ref{specialized analytic}) above].

If we maintain the same value for $\rho$ as in Fig.~\ref{compare}, but
take $\xi \gtrsim 0.03$, the  
curves for $\Lambda_\text{CC}$ and $\Lambda_\text{ML}^\text{NG}$ cannot be distinguished from each other.  
We find in general that for \emph{any} values of $N$, $\sigma_1$, $\sigma_2$, and $\rho$, 
as $\xi \rightarrow 1$, the false dismissal versus false alarm curves for $\Lambda_\text{CC}$ 
and $\Lambda_\text{ML}^\text{NG}$ cannot be distinguished from each other.
Thus, the two statistics are nearly equivalent for Gaussian
signals, as expected.  However, for $\xi \ll 1$, 
Fig.\ \ref{compare} demonstrates that $\Lambda_\text{ML}^\text{NG}$ performs noticeably better than
$\Lambda_\text{CC}$.

We now discuss a comparison of the two statistics in terms of the
minimum gravitational wave energy density necessary for detection,
instead of in terms of the false dismissal versus false alarm curves.  
For a stochastic background with rms strain amplitude 
$h_\text{rms}$, we have $\Omega \propto h^2_\text{rms}$ \cite{Allen
Review}, where $\Omega$ is the gravitational wave energy density.  
For our model signal (\ref{signal model}) we have $h_{\rm rms}^2
\propto \xi \alpha^2$, and comparing this with the 
formula $\rho \propto \xi \alpha^2$ from Eq.\ (\ref{rho2}) shows that
we can interpret the signal to 
noise ratio $\rho$ as the energy density in the stochastic background,
even for non-Gaussian signals.

\begin{figure}
\begin{center}
\epsfig{file=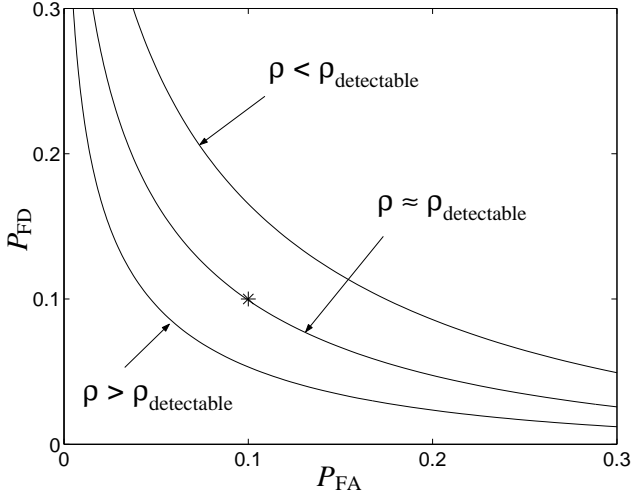,width=8.5cm}
\caption{A family of false dismissal versus false alarm curves for fixed $\xi$.  
Here the detection point, at $P_{\text{FD}*} = P_{\text{FA}*} = 0.1$, is marked with $*$.}
\label{rho detectable}
\end{center}
\end{figure}

\begin{figure}
\begin{center}
\epsfig{file=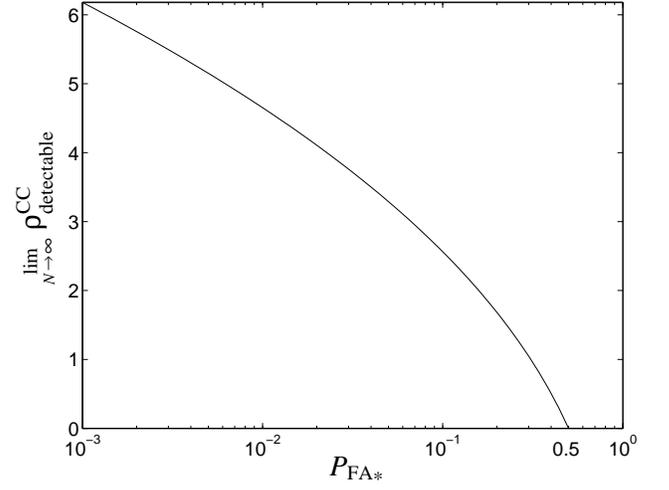,width=8.5cm}
\caption{The minimum detectable signal-to-noise ratio $\rho_\text{detectable}^\text{CC}$ for the cross-correlation statistic $\Lambda_\text{CC}$
as a function of the false alarm probability threshold $P_{\text{FA}*}$. Note that we assume the false dismissal probability
threshold $P_{\text{FD}*} = P_{\text{FA}*}$.}
\label{min rho^G_detectable}
\end{center}
\end{figure}

We compute the minimum detectable energy density or signal-to-noise
ratio $\rho_{\rm detectable}$ as follows.  First, we choose 
thresholds $P_{\text{FA}*}$ and $P_{\text{FD}*}$ for the false alarm
and false dismissal probabilities.  We refer to the pair
$(P_{\text{FA}*},P_{\text{FD}*})$ as the \emph{detection point}. 
For any statistic $\Gamma$, the choice of detection point determines
the detection threshold $\Gamma_*$, and inverting Eq.~(\ref{dependance3})
gives the minimum detectable signal-to-noise ratio 
\begin{equation} 
\rho = \rho_\text{detectable}(P_{\text{FA}*},P_{\text{FD}*},\xi,N),
\end{equation} 
as illustrated in Fig.~\ref{rho detectable}. 
For the cross-correlation statistic $\Lambda_\text{CC}$ the
result is, from Eq.~(\ref{analytic}), 
\begin{eqnarray}
\rho_\text{detectable}^\text{CC} &=& 
\frac{2\sqrt{2}\gamma \left[1 + \gamma\sqrt{2/N}\right]}{ 1 +
  2\gamma^2\left(1 - \frac{3}{\xi} 
  \right)/N } \left[ 1 + O\left( {1 \over \sqrt{N} }\right) \right] \nonumber \\
& & \label{analytic detectable} \\
&=& 2\sqrt{2}\gamma + O\left(\frac{1}{\sqrt{N}}\right) + O\left(
     {\gamma \over \sqrt{N}} \right), \nonumber \\
& & \label{specialized detectable}
\end{eqnarray} 
where $\gamma = \erfc^{-1}(2P_{\text{FA}*})$ and we have assumed that
$P_{\text{FA}*}=P_{\text{FD}*}$.
This relation is plotted in Fig.~\ref{min rho^G_detectable}.

From the results of our simulations, we determine 
$\rho_\text{detectable}(P_{\text{FA}*},P_{\text{FD}*},\xi,N)$ by
numerically solving the equation
\begin{equation} \label{root}
P_\text{FD}(P_{\text{FA}*},\xi,\rho,N) - P_{\text{FD}*} = 0
\end{equation} 
for $\rho$.
Unfortunately, evaluating the function on the left hand side of
Eq.~(\ref{root}) 
is computationally expensive.  Each evaluation involves simulating the
false dismissal versus false alarm curve which 
is itself a computationally intensive task.  Moreover, 
it is only feasible for us to solve Eq.~(\ref{root}) for values of $N$
$\alt 10^4$ while 
a realistic detection scenario for ground based detectors would
involve a year's worth of data sampled at $\sim 100$ 
Hz for which $N\sim 10^9$.  Therefore our conclusions about the
applicability of the method to ground based detectors are based on our
analytic results, as discussed in the Introduction.

\begin{figure}
\begin{center}
\epsfig{file=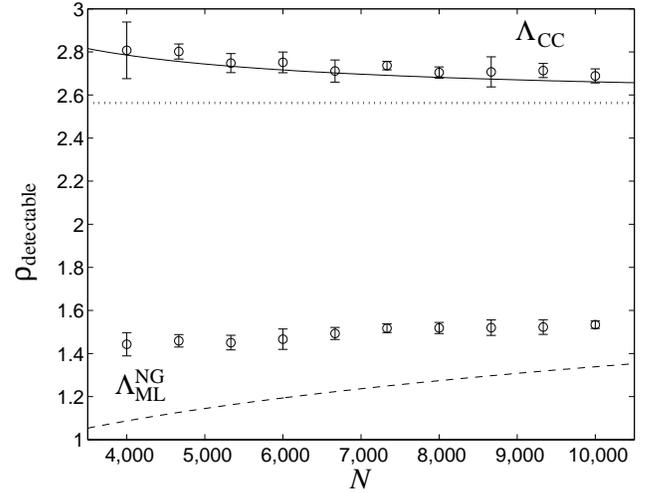,width=8.5cm}
\caption{The minimum detectable signal strength
  $\rho_\text{detectable}$ as a function of
the number of data points $N$,
for the false alarm probability threshold $P_{\text{FA}*}=0.1$, false
dismissal probability threshold $P_{\text{FD}*} = 0.1$, and
Gaussianity parameter $\xi = 0.02$.
The circles are the simulation results, and the error bars 
are estimated from ten
different runs.  The solid curve is the analytical prediction (\ref{analytic detectable}) for 
$\Lambda_\text{CC}$, and the dotted line is the $N \to \infty$ limit
  (\ref{specialized detectable}).
The dashed line is the analytic prediction for $\Lambda_{\rm
  ML}^{\rm NG}$ given by Eqs.\ (\ref{eq:ansA}) and (\ref{eq:ansB}).
}
\label{detection curves}
\end{center}
\end{figure}

\begin{figure}
\begin{center}
\epsfig{file=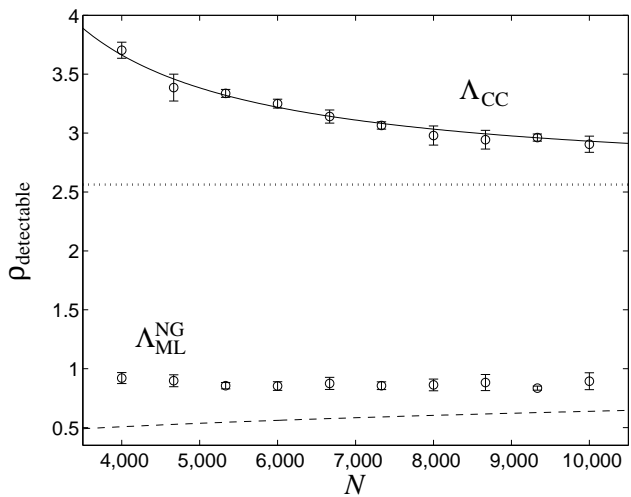,width=8.5cm}
\caption{Same as Fig.\ \protect{\ref{detection curves}} but with $\xi
  = 4.3 \times 10^{-3}$.
}
\label{LargeN}
\end{center}
\end{figure}

Figure \ref{detection curves} shows the results obtained from
numerically solving Eq.~(\ref{root}) for $\rho_{\rm detectable}$ for
the parameter values $\xi = 0.02$, $P_{\text{FA}*} = P_{\text{FD}*} = 0.1$,
and Fig.~\ref{LargeN} shows the corresponding results for $\xi = 4.3
\times 10^{-3}$.  For the cross-correlation statistic, the results are
in good agreement with the analytic prediction 
(\ref{analytic detectable}).

Figure \ref{omega gain} shows the minimum detectable energy density 
as a function of the Gaussianity parameter $\xi$ for $N=10^4$ (corresponding to space based detectors), for the
cross-correlation and maximum likelihood statistics and also for the
burst statistic (\ref{eq:lambdaBdef}).  We again use the values   
$P_{\text{FA}*} =  P_{\text{FD}*} = 0.1$.  The figure shows that the
maximum likelihood statistic performs better than the other statistics
by a factor which is roughly 3 for $\xi$ of order 1\%.  For smaller values
of $\xi$, the maximum likelihood performs increasingly better than the 
cross-correlation statistic, but is eventually comparable to the burst statistic.
Thus the maximum likelihood statistic gives an improvement in sensitivity to backgrounds
composed of roughly $10$ to $10^{3}$ events per year.

Figure \ref{fig:theoretical} is a similar plot, without the Monte Carlo simulation results,
for $N = 10^9$ (corresponding to ground based detectors). Here we use $P_{\text{FA}*} =  P_{\text{FD}*} = 0.01$.
The results are similar to those in Fig.~\ref{omega gain}, except that here the gain in sensitivity
occurs in the band $10^{-5} < \xi < 10^{-3}$.  This band corresponds to $10^4$-$10^6$ events per year.

\subsection{Parameter estimation}
\label{ss:Results for parameter estimation}

\begin{figure}
\begin{center}
\epsfig{file=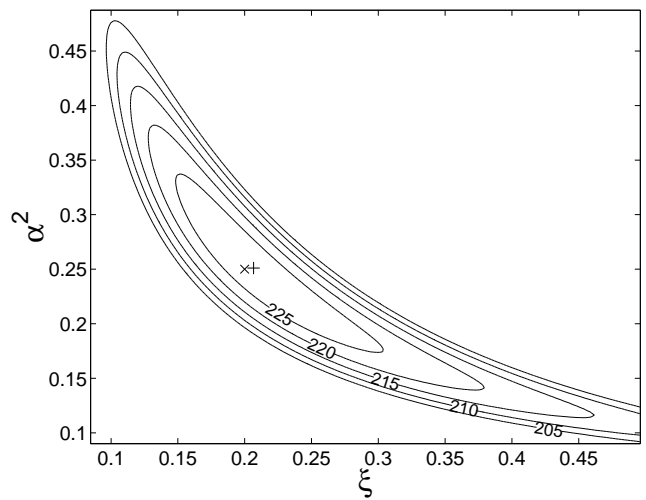,width=8.5cm}
\caption{
Representative contours of $\ln \lambda(\xi,\alpha^2,\hat \sigma_1^2,\hat \sigma_2^2)$.
Here $\rho= 20$ and $N = 1.6 \times 10^5$.  The simulated signal is characterized by $\xi = 0.2$ and 
$\alpha^2 = 0.25$, marked with an $\times$.  The noise is characterized by $\sigma_1^2 = \sigma_2^2 = 1$.
The maximum, marked with a $+$, is found at $\ln \lambda(0.207, 0.251, 0.993, 0.993) = 229$, 
while $\ln \lambda(0.2,0.25,1,1) = 227$.  
}
\label{contours}
\end{center}
\end{figure}

The computation of the maximum likelihood statistic also serves to
measure the parameters of the signal.
The statistic $\Lambda_\text{ML}^\text{NG}$, from Eq.~(\ref{main result2}), can be written as
\begin{equation} \label{main result simple form}
\Lambda_\text{ML}^\text{NG} = \max_{0<\xi\le 1}~ \max_{\alpha^2>0}~ 
			\max_{\sigma^2_1 \ge 0}~ \max_{\sigma^2_2 \ge 0}~ 
			\lambda(\xi, \alpha^2, \sigma^2_1,\sigma^2_2).
\end{equation} 
The point $(\hat \xi,\hat \alpha^2,\hat \sigma_1^2,\hat \sigma^2_2)$ where this maximum is achieved 
is the maximum likelihood estimator for $(\xi,\alpha^2,\sigma_1^2,\sigma^2_2)$. In Fig.~\ref{contours} 
we show contours of the function $\ln \lambda$ for a strong ($\rho = 20$) signal.
This figure shows that both $\xi$ and $\alpha^2$ can be measured with
good accuracy. 

Note that the main benefit of using $\Lambda_{\rm ML}^{\rm NG}$
is that it allows  
us to detect signals that are too weak to be seen using $\Lambda_\text{CC}$.  Using 
$\Lambda_\text{ML}^\text{NG}$ also allows one to test if a detected signal is Gaussian, as obtained 
above, but this is not the main benefit of the method, as there are other, simpler, methods to test 
for non-Gaussianity.

\section{Conclusions}
\label{s:Conclusions}

The use of our maximum likelihood statistic in searches for a
non-Gaussian background gives a gain in sensitivity over the
standard cross-correlation statistic.  Figures \ref{omega gain} and \ref{fig:theoretical} show
that the gain factor can be significant for sufficiently non-Gaussian signals.
However, computing the maximum likelihood statistic requires significantly more
computational power than the cross-correlation statistic.

The analysis presented here must be generalized in several ways before
being usable in gravitational wave detectors.  These generalizations,
listed in order of importance, are:

\begin{itemize}

\item Our signal model (\ref{signal model}) assumes a Gaussian
distribution of amplitudes of the burst events.  This assumption simplified
our analysis and resulted in a statistic with the useful property of 
being nearly equivalent to the cross-correlation statistic in the Gaussian 
signal limit.  In practice however, the distribution of the events
should instead be based on the candidate sources.  For
example, a popcorn-like stochastic background produced by a spatially
uniform distribution 
of standard-candle sources out to some maximum redshift would have a
signal distribution of the form (\ref{signal model}) with the Gaussian
term replaced by a term proportional to $s^{-4}\theta(s-s_{\min})$,
where $\theta$ is the step function and $s_{\rm min}$ is a cutoff
signal strength.

\item One should allow the burst durations to be longer than the
detector resolution time.  For this situation one possibility would
be to preprocess the data with a lowpass filter, and then apply
the techniques developed here.  Another possibility would be to try to
combine the analysis of this paper with the excess power detection
method of Ref.\ \cite{excess power}.

\item Real detector noise always contains non-Gaussian components, so
one needs to generalize the analysis to allow for this.  Such a
generalization for a Gaussian stochastic background can be found in
Refs.\ \cite{robust gaussian,robust gaussian II}.

\item It would be useful to consider a more general signal model which
consists of a superposition of a Gaussian background and a
non-Gaussian background, since the true gravitational wave background
might consist of such a superposition.

\item The analysis needs to be generalized to allow for colored
detector noise, and separated, misaligned detectors.  This
generalization should be fairly straightforward.

\end{itemize}

\begin{acknowledgments}
We thank Wolfgang Tichy, Tom Loredo, Teviet Creighton, and Bernard Whiting for helpful
discussions, and the web site {\it google.com} for providing useful
references on the generalized central limit theorem.
The analytic computations in Appendix \ref{s:appendixC}
were carried out using the software package {\it Mathematica}.
This work was supported in part by National Science
Foundation awards PHY-9722189 and PHY-0140209, the Alfred P. Sloan
foundation, the Radcliffe Institute for Advanced Study, and the
NASA/New York Space Grant Consortium.  
\end{acknowledgments}

\appendix

\section{General form of the likelihood ratio}
\label{s:appendixA}

In this appendix we give two derivations of the general formula
(\ref{general likelihood ratio}) for the likelihood ratio. 
The first derivation is based on Eq.~(\ref{def1}) while the second is based on Eq.~(\ref{def2}).  
We also derive the formula (\ref{distribution relation}) for the posterior probability density 
$p^{(1)}_{\mathcal{V}_s,\mathcal{V}_n|\mathcal{T}}({\bf v}_s,{\bf v}_n|1)$.

\subsection{First derivation}
\label{ss:First derivation}

We can derive Eq.~(\ref{general likelihood ratio}) by using the total probability theorem to 
expand the distributions in the numerator and denominator of Eq.~(\ref{def1}). Note that all 
distributions in this derivation are priors.  

First expand $p_\mathcal{H}(h)$ just in terms of the random variable
$\mathcal{T}$ 
\begin{equation} \label{expansion goal} 
p_\mathcal{H}(h) = P_\mathcal{T}(1)p_\mathcal{H|T}(h|1) 
                 + P_\mathcal{T}(0) p_\mathcal{H|T}(h|0).
\end{equation} 
Expanding $p_\mathcal{H}(h)$ in terms of all the degrees of freedom yields
\begin{eqnarray} 
&& p_\mathcal{H}(h) = \sum_{t=0}^1~ \int_{\Theta_s} d^{Q_s}v_s~ \int d^{ND}s~ 
		\int_{\Theta_n} d^{Q_n}v_n \label{expand1} \\
&&		~\times~  p_{\mathcal{H|T,V}_s,\mathcal{S,V}_n}(h|t,{\bf v}_s,s,{\bf v}_n) 
			  p_{\mathcal{T,V}_s,\mathcal{S,V}_n}(t,{\bf v}_s,s,{\bf v}_n), \nonumber.
\end{eqnarray} 
The ratio of the coefficients of $P_\mathcal{T}(1)$ and $P_\mathcal{T}(0)$ in Eq.~(\ref{expand1}) will 
give the general expression for the likelihood ratio by Eq.~(\ref{def1}).

The conditional distribution for $\mathcal{H}$ in Eq.~(\ref{expand1}) can be translated into a conditional distribution 
for $\mathcal{N}$.  From Eq.~(\ref{detector output matrices}) it
follows that
\begin{equation} \label{trick1}
p_\mathcal{H|S}(h|s)=p_\mathcal{N+S|S}(h|s) = p_\mathcal{N|S}(h-s|s),
\end{equation}
and since $\mathcal{S}$ and $\mathcal{N}$ are statistically independent we obtain
\begin{equation} \label{trick3}
P_\mathcal{H|S}(h|s)=P_\mathcal{N}(h-s).
\end{equation}
Generalizing this argument gives
\begin{equation}  \label{simplify1}
p_{\mathcal{H|T,V}_s,\mathcal{S,V}_n}(h|t,{\bf v}_s,s,{\bf v}_n) = 
p_{\mathcal{N|V}_n}(h-s|{\bf v}_n),
\end{equation} 
since a priori $\mathcal{T}$, $\mathcal{V}_s$, and $\mathcal{S}$ are statistically independent 
of $\mathcal{N}$ and  $\mathcal{V}_n$. For the same reason we can
write the joint distribution that appears in Eq.~(\ref{expand1}) as
\begin{equation} \label{simplify2}
p_{\mathcal{T,V}_s,\mathcal{S,V}_n}(t,{\bf v}_s,s,{\bf v}_n) = 
p_{\mathcal{T,V}_s,\mathcal{S}}(t,{\bf v}_s,s)p_{\mathcal{V}_n}({\bf v}_n).
\end{equation} 

Substituting Eqs.~(\ref{simplify1}) and (\ref{simplify2}) into Eq.~(\ref{expand1}) yields
\begin{eqnarray} 
p_\mathcal{H}(h) &=& \sum_{t=0}^1~ \int_{\Theta_s} d^{Q_s}v_s~ \int d^{ND}s~ 
		\int_{\Theta_n} d^{Q_n}v_n \label{expand2} \\
                &\times&  p_{\mathcal{N|V}_n}(h-s|{\bf v}_n) p_{\mathcal{T,V}_s,\mathcal{S}}(t,{\bf v}_s,s)
		p_{\mathcal{V}_n}({\bf v}_n) .\nonumber
\end{eqnarray}
We can also rewrite the distribution
$p_{\mathcal{T,V}_s,\mathcal{S}}(t,{\bf v}_s,s)$ as
\begin{equation} \label{expand3} 
p_{\mathcal{T,V}_s,\mathcal{S}}(t,{\bf v}_s,s) = p_{\mathcal{S|V}_s,\mathcal{T}}(s|{\bf v}_s,t) 
p_{\mathcal{V}_s|T}({\bf v}_s,t) P_\mathcal{T}(t),
\end{equation} 
by Eq.~(\ref{conditional joint}).
Substituting Eq.~(\ref{expand3}) into Eq.~(\ref{expand2}) and explicitly evaluating the sum over $t$ yields
\begin{widetext}
\begin{eqnarray} 
p_\mathcal{H}(h) &=& P_\mathcal{T}(1)\int_{\Theta_{s1}}d^{Q_s}v_s~ \int d^{ND}s~ \int_{\Theta_n} d^{Q_n}v_n~
		 p_{\mathcal{N|V}_n}(h-s|{\bf v}_n) 
		 p_{\mathcal{V}_n}({\bf v}_n)
		 p_{\mathcal{S|V}_s,\mathcal{T}}(s|{\bf v}_s,1) 
		 p_{\mathcal{V}_s|\mathcal{T}}({\bf v_s}|1) \nonumber \\
		 &+& P_\mathcal{T}(0)\int_{\Theta_n} d^{Q_n}v_n~ 
		 p_{\mathcal{N|V}_s}(h|{\bf v}_n). \label{expand4}
\end{eqnarray}
\end{widetext}
Here we have used the following relations:
\begin{eqnarray} 
p_{\mathcal{S|V}_s,\mathcal{T}}(s|{\bf v}_s\in \Theta_{s1},0) &=& \delta^{ND}(s) \\
p_{\mathcal{V}_s|\mathcal{T}}({\bf v_s}\in \Theta_{s0}|1) &=& 0  \\
p_{\mathcal{V}_s|\mathcal{T}}({\bf v_s}\in \Theta_{s1}|0) &=& 0  \\
\int_{\Theta_{s0}} d^{Q_s}v_s~ p_{\mathcal{V}_s|\mathcal{T}}({\bf v}_s|0) &=& 1.
\end{eqnarray}
By comparing Eqs.~(\ref{expansion goal}) and (\ref{expand4}) we can read off the distributions 
$p_\mathcal{H|T}(h|t)$ and construct Eq.~(\ref{general likelihood
ratio}) from Eq.~(\ref{def1}).  Note that the expression (\ref{def1})
is independent of the space $\Theta_{s0}$ of signal parameters
corresponding to ``no signal present''.

\subsection{Second derivation}
\label{Second derivation}

Here we derive Eq.~(\ref{general likelihood ratio}), and also Eq.~(\ref{distribution relation}), from Eq.~(\ref{def2}).
Consider the distribution 
\begin{equation} \label{simple1} 
p_{\mathcal{T,V}_s,\mathcal{V}_n|\mathcal{H}}(1,{\bf v}_s,{\bf v}_n|h) = 
	\frac{ p_{\mathcal{T,V}_s,\mathcal{V}_n,\mathcal{H}}(1,{\bf v}_s,{\bf v}_n,h) }{ p_\mathcal{H}(h) }.
\end{equation} 
We will justify Eq.~(\ref{general likelihood ratio}) by the defining relation Eq.~(\ref{def2}), which explicitly refers 
to priors and posteriors.  Therefore we now append the appropriate superscripts as bookkeeping devices.  
Eq.~(\ref{simple1}) then reads
\begin{equation} \label{simple2} 
p^{(1)}_{\mathcal{T,V}_s,\mathcal{V}_n}(1,{\bf v}_s,{\bf v}_n) = 
\frac{ p^{(0)}_{\mathcal{T,V}_s,\mathcal{V}_n,\mathcal{H}}(1,{\bf v}_s,{\bf v}_n,h) }
     { p^{(0)}_\mathcal{H}(h) }.
\end{equation}

Using the expansion of $p_\mathcal{H}(h)$ given by Eq.~(\ref{expand4}), and what we will justify is the 
likelihood ratio $\Lambda$ given by Eq.~(\ref{general likelihood ratio}), we have
\begin{equation} \label{simple3} 
p^{(1)}_{\mathcal{T,V}_s,\mathcal{V}_n}(1,{\bf v}_s,{\bf v}_n) = 
\frac{ \left[ \frac{p^{(0)}_{\mathcal{T,V}_s,\mathcal{V}_n,\mathcal{H}}(1,{\bf v}_s,{\bf v}_n,h)}
			{\int_{\Theta_n} d^{Q_n}v_n'~ p^{(0)}_{\mathcal{H|V}_n}(h|{\bf v}_n)
			p^{(0)}_{\mathcal{V}_n}({\bf v}_n)} \right] }
			{ \Lambda P^{(0)} + 1 - P^{(0)} }.
\end{equation} 
Expanding the uppermost numerator in Eq.~(\ref{simple3}) over
$\mathcal{S}$ by the total probability theorem gives
\begin{eqnarray}  \label{simple4}
p^{(0)}_{\mathcal{T,V}_s,\mathcal{V}_n,\mathcal{H}}(1,{\bf v}_s,{\bf v}_n,h) &=& \int d^{ND}s \\
&\times& p^{(0)}_{\mathcal{T,V}_s,\mathcal{V}_n,\mathcal{H,S}}(1,{\bf v}_s,{\bf v_n},h,s), \nonumber
\end{eqnarray} 
and rewriting this gives
\begin{eqnarray} \label{simple expand}
&& p^{(0)}_{\mathcal{T,V}_s,\mathcal{V}_n,\mathcal{H,S}}(1,{\bf v}_s,{\bf v_n},h,s) = 
			p^{(0)}_{\mathcal{N|V}_s}(h-s|{\bf v}_s) \nonumber \\
&& \times		p^{(0)}_{\mathcal{V}_n}({\bf v}_n) 
			p^{(0)}_{\mathcal{S|V}_s,\mathcal{T}}(s|{\bf v}_s,1) 
			p^{(0)}_{\mathcal{V}_s|\mathcal{T}}({\bf v}_s,1)P^{(0)}. 
\end{eqnarray} 
After putting Eq.~(\ref{simple expand}) into Eq.~(\ref{simple4}), substitute the result into Eq.~(\ref{simple3}).
Using $\Lambda({\bf v}_s,{\bf v}_n)$ given by Eq.~(\ref{likelihood function}) then yields
\begin{equation} \label{simple5} 
P^{(1)}p^{(1)}_{\mathcal{V}_s,\mathcal{V}_n|\mathcal{T}}({\bf v}_s,{\bf v}_n|1) = 
\frac{ \Lambda({\bf v}_s,{\bf v}_n)P^{(0)} }{ \Lambda P^{(0)} + 1 - P^{(0)} }.
\end{equation} 
On the left hand side of Eq.~(\ref{simple5}) we have used 
\begin{equation} 
p^{(1)}_{\mathcal{T,V}_s,\mathcal{V}_n}(1,{\bf v}_s,{\bf v}_n) = P^{(1)}
p^{(1)}_{\mathcal{V}_s,\mathcal{V}_n|\mathcal{T}}({\bf v}_s,{\bf v}_n|1).
\end{equation} 

Integrate Eq.~(\ref{simple5}) over $\Theta_n$ and $\Theta_{s1}$ using Eq.~(\ref{likelihood function def}) 
and the normalization requirement 
\begin{equation} \label{simple normalization}
\int_{\Theta_{s1}}d^{Q_s}v_s \int_{\Theta_n}d^{Q_n}v_n~ 
p^{(1)}_{\mathcal{V}_s,\mathcal{V}_n|\mathcal{T}}({\bf v}_s,{\bf v}_n|1) = 1
\end{equation}  
to get 
\begin{equation} \label{p1}
P^{(1)} = \frac{\Lambda P^{(0)}}{\Lambda P^{(0)} + 1 - P^{(0)}}.
\end{equation} 
Use Eq.~(\ref{p1}) and Eq.~(\ref{simple5}) to form the ratio on the left hand side of 
Eq.~(\ref{distribution relation}) .  This justifies Eq.~(\ref{distribution relation}). 

Integrate Eq.~(\ref{distribution relation}) over $\Theta_n$ and $\Theta_{s1}$ using 
Eq.~(\ref{likelihood function def}) and Eq.(~\ref{simple normalization}) to see that the 
defining relation Eq.~(\ref{def2}) is satisfied and thus Eq.~(\ref{general likelihood ratio}) 
is justified.

\medskip

\section{Analytical expressions for false dismissal versus false alarm
curves for cross-correlation statistic}
\label{s:appendixB}
This appendix derives the analytical form (\ref{analytic})
of the false dismissal versus false alarm curves for the
cross-correlation statistic $\Lambda_\text{CC}$ in the large
$N$ limit, for both Gaussian and non-Gaussian signals.  
A derivation for Gaussian signals can be found in Sec.~IV of
Ref.~\cite{Allen Romano}. 

As noted in Sec.~\ref{ss:Gaussian signal}, the statistics
$\Lambda_\text{CC}$ and $\hat\alpha^2$ 
are equivalent in the large $N$ limit. Thus, in this limit, the false
dismissal versus false alarm curves 
can be found by evaluating Eqs.~(\ref{simple Pfa}) and (\ref{simple
Pfd}) with $\Gamma$ replaced by $\hat\alpha^2$.  
The relation (\ref{gaussian estimator1}) between the statistics ${\bar
\alpha}^2$ and ${\hat \alpha}^2$ implies the following relation
between their probability distributions 
$p_{\hat\alpha^2|\mathcal{T}}(x|t)$ and $p_{\bar\alpha^2|\mathcal{T}}(x|t)$:
\begin{equation}
p_{\hat\alpha^2|\mathcal{T}}(x|t)=
\theta(x) p_{\bar\alpha^2|\mathcal{T}}(x|t)
+\delta(x) \, \int_{-\infty}^0
dy \, p_{\bar\alpha^2|\mathcal{T}}(y|t). 
\end{equation}
Inserting this formula into Eqs.\ (\ref{simple Pfa}) and (\ref{simple
Pfd}) gives 
\begin{eqnarray}  
P_\text{FA}(\hat\alpha^2_*) &=&\left\{ 
\begin{array}{ll} 
 \displaystyle \int_{\hat\alpha^2_*}^\infty dx~p_{\bar\alpha^2|\mathcal{T}}(x|0) 
	& \text{ if } \hat\alpha^2_* > 0 \\
 \displaystyle 1  
	& \text{ if } \hat\alpha^2_* \le 0 \\
\end{array} \right. ,\label{Pfa} \\
P_\text{FD}(\hat\alpha^2_*) &=&\left\{ 
\begin{array}{ll} 
 \displaystyle 1 - \int_{\hat\alpha^2_*}^\infty dx~p_{\bar\alpha^2|\mathcal{T}}(x|1) 
	& \text{ if } \hat\alpha^2_* > 0 \\
 \displaystyle 0 
	& \text{ if } \hat\alpha^2_*\le 0 \\
\end{array} \right. \nonumber . \\ && \label{Pfd}
\end{eqnarray}

In the large $N$ limit, the distribution
$p_{\bar\alpha^2_s|\mathcal{T}}(x|t)$ must be Gaussian by the central
limit theorem, and   
therefore this distribution is characterized entirely by its mean $\left<
\bar\alpha^2_t \right>$ and variance  
$[\Delta(\bar\alpha^2_t)]^2$.  From Eqs.\ (\ref{detector output
matrices}), (\ref{assumption4}), (\ref{baralphadef}), (\ref{gaussian
estimator1}) and (\ref{signal model}), these are given by
\begin{eqnarray} 
\left< \bar\alpha^2_0 \right> &=& 0 \label{mean0}\\
\Delta(\bar\alpha^2_0)  &=& \frac{\sigma_1\sigma_2}{\sqrt{N}}  \label{var0} \\
\left< \bar\alpha^2_1 \right> &=& \xi\alpha^2 \label{mean1} \\
\Delta(\bar\alpha^2_1) &=& \sqrt{ \frac{ \xi\alpha^4(3-\xi) + \xi\alpha^2(\sigma_1^2+\sigma_2^2) 
	+ \sigma_1^2 \sigma_2^2} {N} }. 
\nonumber \\ && \label{var1}
\end{eqnarray} 
Substituting Gaussian distributions, with means and variances determined by 
Eqs.~(\ref{mean0})-(\ref{var1}), into Eqs.~(\ref{Pfa}) and (\ref{Pfd}) yields
\begin{widetext}
\begin{eqnarray} 
P_\text{FA}(\hat\alpha^2_*,\sigma_1,\sigma_2,N) &=& 
\left\{\begin{array}{ll} 
 \displaystyle \frac{1}{2} \erfc \left( \frac{\hat\alpha^2_*}{\sigma_1\sigma_2}\sqrt{\frac{N}{2}} \right)  
	& \text{ if } \hat\alpha^2_* > 0 \\
 \displaystyle 1  
	& \text{ if } \hat\alpha^2_* \le 0 \\
\end{array} , \right. \label{analytic1} \\
P_\text{FD}(\hat\alpha^2_*,\xi,\alpha,\sigma_1,\sigma_2,N) &=& 
\left\{\begin{array}{ll}
 \displaystyle 1 - \frac{1}{2} \erfc \left[ \left( \hat\alpha^2_*-\xi\alpha^2 \right)
              \sqrt{ \frac{N}{2 \left[ \xi\alpha^4(3-\xi) + \xi\alpha^2(\sigma_1^2+\sigma_2^2) 
              + \sigma_1^2 \sigma_2^2 \right]} }\right]  
	& \text{ if } \hat\alpha^2_* > 0 \\
 \displaystyle 0  
	& \text{ if } \hat\alpha^2_* \le 0 \\
\end{array} \right. .\nonumber \\ &&  
\label{analytic2}
\end{eqnarray} 
\end{widetext}
If we now eliminate $\hat\alpha^2_*$ between Eqs.~(\ref{analytic1})
and (\ref{analytic2}), change variables from $\alpha$ to $\rho$
using Eq.~(\ref{rho2}), and set $\sigma_1 = \sigma_2$, 
we obtain Eq.~(\ref{analytic}).

\section{Asymptotic behavior of maximum likelihood statistic}
\label{s:appendixC}

In this appendix we derive the large-$N$ behavior of 
the maximum likelihood statistic $\Lambda^{\rm NG}_{\rm ML}$.
From Eq. (\ref{main result2}), we can write the statistic in the form
\begin{equation}
\Lambda_{\rm ML}^{\rm NG}(h) = \exp \left[ N {\cal L}(h) \right]
\label{lambdadef}
\end{equation}
with
\begin{equation}
{\cal L}(h) = \max_{\sigma_1,\sigma_2,\xi,\alpha} \, 
g(\sigma_1,\sigma_2,\xi,\alpha,h)
\label{lambdadef1a}
\end{equation}
where
\begin{equation}
g = \frac{1}{N}
\sum_{k=1}^N \, g_k(\sigma_1,\sigma_2,\xi,\alpha),
\label{lambdadef1}
\end{equation}
and the function $g_k = g_k(\sigma_1,\sigma_2,\xi,\alpha)$ is
given by 
\begin{equation}
e^{g_k} = \xi A_k(\alpha) + (1 - \xi) A_k(0)
\end{equation}
with
\begin{eqnarray}
A_k(\alpha) &=&  
 { \exp \left[ \frac{\left( \frac{h_1^k}{\sigma^2_1} + \frac{h_2^k}{\sigma^2_2}\right)^2}
          {2\left( \frac{1}{\sigma^2_1} + \frac{1}{\sigma^2_2} + \frac{1}{\alpha^2} \right)} 
          - \frac{\left( h_1^k\right)^2}{2\sigma^2_1} - \frac{\left(
	h_2^k\right)^2}{2\sigma^2_2} +1 \right]} \nonumber \\
 && \times { {\bar \sigma}_1 {\bar \sigma}_2  \over 
\sqrt{\sigma^2_1 \sigma^2_2 + \sigma^2_1 \alpha^2 + \sigma^2_2 \alpha^2}}.
\label{f def}
\end{eqnarray}
We denote by ${\tilde \sigma}_1$, ${\tilde \sigma}_2$, ${\tilde \xi}$ and
${\tilde \alpha}$ the ``true'' parameters governing the distribution of
the quantities $h_1^k$ and $h_2^k$ according to
Eqs. (\ref{detector output matrices}), (\ref{assumption2}),
(\ref{assumption4}), and (\ref{signal model}), with untilded
quantities replaced by the corresponding tilded quantities.
[These ``true parameters'' were denoted by $\sigma_1$, $\sigma_2$,
$\xi$ and $\alpha$ in the body of the paper.]
We define ${\tilde \rho}$ to
be the signal-to-noise ratio (\ref{rho2}) with untilded
quantities replaced by tilded quantities:
\begin{equation}
{\tilde \rho} \equiv \frac{ {\tilde \xi} {\tilde \alpha}^2 \sqrt{N} }{ {\tilde
\sigma}_1 {\tilde \sigma}_2}.  
\label{rho2bar}
\end{equation}
For simplicity, in this appendix we restrict attention to the case
${\tilde \sigma}_1 = {\tilde \sigma}_2$.  Then, without loss of
generality, we can take ${\tilde \sigma}_1 = {\tilde \sigma}_2 =1$ by
rescaling our units of strain amplitude.

We discuss separately the computation of the false alarm and false
dismissal probabilities, as different techniques are required to
compute each.

\subsection{False dismissal probability}

The false dismissal probability for the statistic (\ref{lambdadef1a})
will be some function
\begin{equation}
P_{\rm FD} = P_{\rm FD}({\cal L}_*, N, {\tilde \xi}, {\tilde \rho})
\end{equation}
of the threshold ${\cal L}_*$ on ${\cal L}$, the number of data points
$N$, the Gaussianity parameter ${\tilde \xi}$ and signal-to-noise
ratio ${\tilde \rho}$ of the signal.  For applications to ground based
detectors, we will have ${\tilde \rho} \sim $ (a few), in order that
the signal be detectable, $N \sim 10^9$, and $10^{-3} \alt {\tilde
\xi}\le 1$.  Therefore it would be useful to find approximate analytic
expressions for the false alarm probability in the limit of large
$N$.  There are actually several different, large $N$ regimes in the
three dimensional parameter space $(N, {\tilde \xi}, {\tilde \rho})$
that one might explore: 
\begin{itemize}
\item The limit $N \to \infty$ with ${\tilde \alpha}$ and ${\tilde
\xi}$ held fixed.  This corresponds to fixing the stochastic
background signal and going to a limit of long observation times.  In
this limit we have ${\tilde \rho} \propto \sqrt{N}$ which diverges.
This is not a very realistic limit to explore.

\item The limit $N \to \infty$ with ${\tilde \rho}$ and ${\tilde \xi}$
held fixed.  In this limit, the signal-to-noise ratio is held fixed,
and correspondingly the amplitude ${\tilde \alpha}$ of the stochastic
background signal goes to zero, from Eq.\ (\ref{rho2bar}).  This would
be the most natural 
limit to explore.  However, in this limit the statistical error
$\Delta {\tilde \xi}$ in our measurement of the Gaussianity parameter
would diverge, from Eq.\ (\ref{Deltaxi}), and therefore in this limit
we do not expect to be able to compute analytically the value of the
parameter $\xi$ which achieves the maximum in Eq.\ (\ref{lambdadef1a}).
The analytic approximation methods which we discuss below do not work in this
regime.  [In addition our Monte Carlo simulations show that the maximum
likelihood statistic itself does not perform any better than the
cross-correlation statistic in this regime, as discussed in the
Introduction.]

\item The limit we actually explore is the limit $N \to \infty$ with
${\tilde \xi}$ fixed and ${\tilde \rho}$ scaling $\propto N^{1/4}$,
corresponding to ${\tilde \alpha} \propto N^{-1/8}$.  The reason for
our choosing to explore this particular limit is simply that it is
amenable to analytic computations.  Fractional corrections to our
analytic results should scale like $1/N$ or as $1 / {\tilde \rho}^4$.  
Since ${\tilde \rho} \sim $ (a few) at the threshold for detection, the
approximation should be good to $10\% - 20\%$ or so.

\end{itemize}

We now turn to a discussion of the computational technique.  
We write 
\begin{equation}
{\tilde \alpha} = {\tilde \alpha}_0 N^{-1/8},
\label{alpha0def}
\end{equation}
where ${\tilde \alpha}_0$ is independent of $N$.
Correspondingly, from Eq.\ (\ref{signal model}) we can write
\begin{equation}
s^k = N^{-1/8} {\hat s}^k,
\end{equation}
where the distribution of ${\hat s}^k$ is given by Eq.\ (\ref{signal
model}) with $\xi$ replaced by ${\tilde \xi}$ and $\alpha$ replaced
by ${\tilde \alpha}_0$.  In particular, the distribution of ${\hat
s}^k$ is independent of $N$.
In computing the maximum over $(\xi,\alpha,\sigma_1, \sigma_2)$ in
Eq.\ (\ref{lambdadef1a}), it is useful 
change variables from $\alpha$ to $\kappa$ defined by
\begin{equation}
\kappa = \rho N^{-1/4} = {\xi \alpha^2 N^{1/4}  \over \sigma_1
  \sigma_2},
\label{kappadef}
\end{equation}
which we expect to be independent of $N$ to leading order in the large
$N$ limit.  The value of the variable $\kappa$ that characterizes the
signal is 
\begin{equation}
{\tilde \kappa} = {\tilde \rho} N^{-1/4} = {{\tilde \xi} {\tilde
    \alpha}_0^2  \over {\tilde 
    \sigma}_1 {\tilde \sigma}_2};
\end{equation}
cf.\ Eqs.\ (\ref{alpha0def}) and (\ref{kappadef}).

We now consider fixed realizations of the infinite sequences of random
variables $n_1^k$, $n_2^k$ and ${\hat s}^k$, and 
$1 \le k < \infty$, and examine the limiting
behavior of ${\cal L}(h)$ as $N \to \infty$.   
We compute this limiting behavior by substituting into 
the right hand side of Eq. (\ref{lambdadef})
the relations 
\begin{equation}
h_1^k = n_1^k + N^{-1/8} {\hat s}^k \ \ \ \ \ 
h_2^k = n_2^k + N^{-1/8} {\hat s}^k,
\end{equation}
writing $\alpha$ in terms of $\kappa$ using Eq.\ (\ref{kappadef}), and
expanding in powers of $N^{-1/8}$.
The result is an expression which can be written in terms of 
the sums $Q_{abc}$ defined by 
\begin{equation}
\label{sums0}
Q_{abc} = \frac{1}{N}\sum_{k=1}^N \left( \hat s^k \right)^a 
				  \left( n_1^k    \right)^b 
				  \left( n_2^k    \right)^c,
\end{equation}
where $a$, $b$, and $c$ are non-negative integers.
From the central limit theorem we can write
\begin{equation}\label{sums}
Q_{abc} = \mu_{abc} + \frac{1}{\sqrt{N}} \Delta_{abc},
\end{equation}
where $\mu_{abc}=\left< Q_{abc} \right>$ are computable functions of
${\tilde \xi}$ and ${\tilde \alpha}$, and where the random variables 
$(\Delta_{100},\Delta_{010},\ldots)$
converge in distribution 
% FOOTNOTE %%%%%%%%%%%%%%%%%%%%%%%%%%%%%%%%%%%%%%%%%%%%%%%%%%%%%%%%%%%%%%%%%%%%%%%%%%%%%%%%%%%%%%%%%%%%%%%%%%
\footnote{See chapter 8 of Ref.\ \cite{Papoulis} for definitions of different
notions of convergence for sequences of random variables. }
%%%%%%%%%%%%%%%%%%%%%%%%%%%%%%%%%%%%%%%%%%%%%%%%%%%%%%%%%%%%%%%%%%%%%%%%%%%%%%%%%%%%%%%%%%%%%%%%%%%%%%%%%%%%%
as $N \rightarrow \infty$ to a multivariate Gaussian of zero mean whose variance-covariance 
matrix is independent of $N$.  Thus, in particular the joint distribution of all 
$\Delta_{abc}$'s is $N$-independent in limit that $N \rightarrow \infty$.

We define the vector
\begin{equation} 
{\bf v} = (v^1,v^2,v^3,v^4) = (\xi,\kappa,\sigma_1^2, \sigma_2^2),
\end{equation} 
We denote the value of ${\bf v}$ that achieves the maximum in Eq.\
(\ref{lambdadef1a}) by $\hat{{\bf v}}$:
\begin{equation} 
g( \hat{{\bf v}}) = \max_{{\bf v}}~g({\bf v}),
\end{equation} 
where $\hat{{\bf v}} = (\hat \xi, \hat \kappa,\widehat{\sigma_1^2},\widehat{\sigma_2^2})$.
These estimators satisfy a system of four equations \footnote{Here we are assuming that the maximum is achieved as a local maximum in the interior of the 4 dimensional parameter space.  Cases when the maximum is achieved on the boundary are discussed below.}
\begin{equation} \label{hard eq system}
\left. \frac{\partial g }{\partial v^l} \right|_{{\bf v} = \hat{{\bf v}}} = 0.
\end{equation} 
We solve Eq.~(\ref{hard eq system}) perturbatively.  
First assume that the estimators can be expanded in the form
\begin{equation} \label{assume expand}
\widehat{v^l} = \sum_{j=0}^{\infty} \widehat{v^{l}}^{[j]} \epsilon^j,
\end{equation} 
where for ease of notation we have defined
$\epsilon = N^{-1/8}$.  
We define the expansion coefficients $v^{l[j]}$ analogously by an
expansion of the form (\ref{assume expand}) 
but without the hats.  Now using Eq.\ (\ref{sums}) the function $g$
can be expanded as a power 
series in $\epsilon$ whose coefficients are functions of
$v^{l[k]}$, $\mu_{abc}$, and $\Delta_{abc}$:
\begin{equation} 
g({\bf v}) = \sum_{j=0}^{\infty} 
g^{[j]}\left[ v^{l[k]}, \mu_{abc},\Delta_{abc} \right] \epsilon^j.
\label{gexpand}
\end{equation} 
Substituting the expansions (\ref{assume expand}) and (\ref{gexpand}) into
the condition 
(\ref{hard eq system}) for a local extremum 
gives an infinite set of equations which must collectively be satisfied by 
the coefficients $\widehat{v^{l}}^{[j]}$
\begin{equation} \label{easy set}
\left. \frac{\partial g^{[j]}}{\partial v^{l[k]}} \right|_{v^{m[n]} = \widehat{v^m}^{[n]}} = 0.
\end{equation}
We solve these equations order by order to determine the coefficients 
$\widehat{{v^l}}^{[j]}$, and thereby justify a posteriori the ansatz
(\ref{assume expand}).

We find that in order to compute the leading order expression for
${\cal L}$, we must obtain the expansion for ${\hat \xi}$ to zeroth
order in $\epsilon$, the expansion for ${\hat \kappa}$ to fourth order
in $\epsilon$, and the expansions of ${\hat {\sigma_1^2}}$ and ${\hat
{\sigma_2^2}}$ to sixth order in $\epsilon$.  
The leading order results are
\begin{eqnarray}
\label{kappaans}
{\hat \kappa} &=& {\tilde \kappa} + \epsilon^2 X + O(\epsilon^3), \\
\label{xians}
{1 \over {\hat \xi}} &=& {1 \over {\tilde \xi}} + {Y \over \sqrt{6}
  {\tilde \kappa}^2} + O(\epsilon), \\
\widehat{\sigma_1^2} &=& 1 + O(\epsilon^2), \\
\widehat{\sigma_2^2} &=& 1 + O(\epsilon^2),
\label{sigmaans}
\end{eqnarray}
where
\begin{eqnarray}
\label{eq:Xdef}
X = \Delta_{011}  
\end{eqnarray}
and
\begin{eqnarray}
\label{eq:Ydef}
 Y &=& {1 \over 8 \sqrt{6}} \bigg[ 4 (\Delta_{031} + \Delta_{013}) - 12
(\Delta_{002} + \Delta_{020}) \nonumber \\
&& - 24 \Delta_{011} + \Delta_{040} + \Delta_{004} + 6 \Delta_{022} \bigg].
\end{eqnarray}
Using Eqs.\ (\ref{assumption2}), (\ref{signal model}), (\ref{sums0}) and
(\ref{sums}) one can show that the random variables $X$ and $Y$ are
independent Gaussian random variables of zero mean and unit variance.

In deriving Eqs.\ (\ref{kappaans}) -- (\ref{sigmaans}) we assumed that
the value of ${\bf v}$ which achieves the maximum in Eq.\
(\ref{lambdadef1a}) corresponds a local maximum.  However, if the
right hand side of Eq.\ (\ref{kappaans}) is negative, the maximum will
instead be achieved on the boundary of the parameter space at ${\hat
\kappa} =0$, since the variable $\kappa$ must be non-negative.
Similarly, if the right hand side of Eq.\ (\ref{xians}) is less than
1, the maximum will be achieved at ${\hat \xi} =1$, since $1/\xi$ must
lie in the interval $[1,\infty)$.

Substituting the results (\ref{kappaans}) -- (\ref{sigmaans}) [together
with the higher order corrections to those results which we have not
shown] into the expansion for the statistic ${\cal L}$, and taking
into account the various special cases discussed in the last
paragraph, gives
\begin{eqnarray}
{\cal L} &=&  \bigg[
{1 \over 2} \left(Y + \sqrt{6} q {\tilde \kappa}^2 \right)^2 \epsilon^8 
\, \theta
\left(Y + \sqrt{6} q {\tilde \kappa}^2 \right) \nonumber \\
&& + {1 \over 2} ({\tilde \kappa} + \epsilon^2 X)^2 \epsilon^4 
- {\tilde \kappa}^3 \epsilon^6 + {7 \over 4} {\tilde \kappa}^4
\epsilon^8 \nonumber \\
 && 
+ {\tilde \kappa} U \epsilon^7 + {\tilde \kappa} V \epsilon^8
\bigg] \theta({\tilde \kappa} + \epsilon^2 X) + O(\epsilon^9).
\label{eq:ML1}
\end{eqnarray}
Here $\theta(x)$ is the step function and
\begin{eqnarray}
\label{eq:qdef}
q &=& {1 \over {\tilde \xi}}-1, \\
U &=& \Delta_{101} + \Delta_{110}, \\
V &=& \Delta_{200} - {1 \over 2} {\tilde \kappa} (\Delta_{002} +
\Delta_{020}) - 2 {\tilde \kappa} \Delta_{011}.
\end{eqnarray}
We note that the corresponding expression for the statistic
$(\ln \Lambda_{\rm ML}^{\rm G})/N$ [which is equivalent to the
cross-correlation statistic by Eq.\ (\ref{Gaussian statistic})] is
given by Eq.\ (\ref{eq:ML1}) with the first term in the square
brackets dropped.

Next we drop all the
terms in the square bracket in Eq.\ (\ref{eq:ML1}) other than the
first two terms.  The 
reason is that these terms will give corrections that are smaller than
the terms retained (both in expected value and in fluctuations) by a
factor of
$$
{\tilde \kappa} \epsilon^2 = {{\tilde \rho} \over \sqrt{N}},
$$
which will be small compared to unity for all cases we are interested
in.  This gives for the false dismissal probability the expression
\begin{eqnarray}
P_\text{FD} &=& P({\cal L} < {\cal L}_*) \nonumber \\
 &=& \int_{\cal R} {dx dy  \over
2 \pi} \exp \left[ -{(x-x_0)^2 \over 2} - {(y-y_0)^2 \over 2} \right], \nonumber \\ & & 
\end{eqnarray}
where
\begin{eqnarray}
\label{eq:def1}
x_0 &=& {\tilde \kappa}/\epsilon^2 \\
y_0 &=& \sqrt{6} q {\tilde \kappa}^2  \\
r_0 &=& \sqrt{2 N {\cal L}_*}.
\label{eq:def4}
\end{eqnarray}
Here the region ${\cal R}$ in the $x,y$ plane is the union of the
two regions
\begin{eqnarray}
x &\ge& 0 \nonumber \\
y &\ge& 0 \nonumber \\
x^2 +  y^2 &\le& r_0^2
\label{eq:region1a}
\end{eqnarray}
and
\begin{eqnarray}
y &\le& 0 \nonumber \\
0 \le x &\le& r_0.
\label{eq:region2a}
\end{eqnarray}

The integral over the region (\ref{eq:region2a}) is
\begin{equation}
P_{\rm FD}^{(1)} = 
{\cal P}(-y_0)[ {\cal P}(r_0 - x_0) - {\cal P}(-x_0)],
\label{eq:1}
\end{equation}
where 
\begin{equation}
{\cal P}(x) \equiv 1 - {1 \over 2} \erfc (x/\sqrt{2}) = \int_{-\infty}^x
dt {1 \over \sqrt{2 \pi}} \exp[-t^2/2].
\end{equation}
The integral over the region (\ref{eq:region1a}) can be written as
\begin{eqnarray}
P_{\rm FD}^{(2)} &=& {1 \over 2 \pi} \int_0^{\pi/2} d\theta \,
\int_{0}^{r_0} dr \, r 
\nonumber \\ &\times&
\exp \left[ - {1 \over 2} (r \cos \theta -
x_0)^2 - {1 \over 2} (r \sin \theta - y_0)^2 \right]. \nonumber \\ 
& & \label{eq:pFD1}
\end{eqnarray}
The integrand in (\ref{eq:pFD1})
peaks
at $r \cos \theta =  
x_0$, $r \sin \theta = y_0$.  In order for $P_{\rm FD}$ to be
small, its necessary that this peak occurs outside the domain of
integration, at $r > r_0$.  So we must have
\begin{equation}
x_0^2 + y_0^2 \ge r_0^2.
\label{eq:constraint}
\end{equation}
The criterion $x_0 \ge r_0$ is, in order of magnitude, just the usual
criterion for detectability with the cross-correlation statistic.  The
criterion $y_0 \agt r_0$ reduces to, in order of magnitude,
\begin{equation}
\xi \alt {\rho^2 \over \sqrt{N}}
\end{equation}
which is what we claimed earlier to be the regime where the maximum
likelihood statistic starts
to work well, cf.\ Sec.\ \ref{s:MLS} above.  

Evaluating the integral (\ref{eq:pFD1}) using the Laplace
approximation gives
\begin{eqnarray}
P_{\rm FD}^{(2)} &=& {1 \over r_0 (\lambda-1) \sqrt{2 \pi \lambda}}
\exp \left[ 
- {1 \over 2} r_0^2 (\lambda-1)^2 \right] \nonumber \\
 && \times \left[ 1 + O\left({1
\over r_0}\right)\right],
\label{eq:final}
\end{eqnarray}
where we define the variables $\lambda$ and $\gamma$ by
\begin{equation}
(x_0,y_0) = r_0 \lambda (\cos \gamma, \sin \gamma).
\label{eq:lambdadef}
\end{equation}
However, the result (\ref{eq:final}) is not very accurate for small
$r_0$.  Alternatively we can integrate over $r$ in Eq.\
(\ref{eq:pFD1}) to obtain
\begin{widetext}
\begin{eqnarray}
&& P_{\rm FD}^{(2)} = \int_0^{\pi/2} d\theta \left\{ 
\frac{1}{2\pi} e^{\frac{{{r_0}}^2\,\left( 1 + {\lambda }^2 \right) }{2}} 
\left[ e^{\frac{{{r_0}}^2}{2}} - e^{{{r_0}}^2\,\lambda \,\cos (\gamma  - \theta )} \right] \right. \nonumber \\
&& + \left. 
\frac{r_0 \lambda}{2\sqrt{2\pi}} 
e^{\frac{ {{r_0}}^2 \lambda^2 }{4} 
   \left[ \cos (2\,\left\{ \gamma  - \theta  \right\} ) -1 \right] }
\cos \left( \gamma  - \theta \right)
     \left[ \erf\left( \frac{{r_0}\,\lambda \cos \{\gamma  - \theta \}}{{\sqrt{2}}} \right) +
            \erf\left( \frac{{r_0}\,\left\{ 1 - \lambda \,\cos [\gamma  - \theta ]  \right\} }{{\sqrt{2}}} \right) \right] 
\right\}, \label{eq:c}
\end{eqnarray}
\end{widetext}
where 
\begin{equation} 
\erf(x) = \frac{2}{\sqrt{\pi}}\int_0^x dy~ e^{-y^2}.
\end{equation} 
The integral (\ref{eq:c}) can be evaluated numerically.
The false dismissal probability is then given by
\begin{equation}
P_{\rm FD} = P_{\rm FD}^{(1)} + P_{\rm FD}^{(2)},
\label{eq:ansA}
\end{equation}
with $P_{\rm FD}^{(1)}$ given by Eq.\ (\ref{eq:1})
and $P_{\rm FD}^{(2)}$ given by Eq.\ (\ref{eq:c}).

\subsection{False alarm probability}

The false alarm probability is some function
\begin{equation}
P_{\rm FA} = P_{\rm FA}({\cal L}_*,N)
\end{equation}
of the threshold ${\cal L}_*$ value of the detection statistic
(\ref{lambdadef1a}) and of the number of data points $N$.  It does not
depend on the signal parameters ${\tilde \rho}$ and ${\tilde \xi}$
because no signal is present.  We would like to evaluate this quantity
in the large $N$ limit.

We start by rewriting the statistic (\ref{lambdadef1a}) in the form
\begin{equation}
{\cal L} = \max_{\bf v} \left\{ {1 \over N} \sum_{k=1}^N \ln A_k(0) 
+ {1 \over N} \sum_{k=1}^N \ln \left[ 1 + \xi {\cal D}_k(\alpha)
  \right] \right\},  
\label{lambdadef4}
\end{equation}
where
\begin{equation}
{\cal D}_k(\alpha) = { A_k(\alpha) \over A_k(0)} -1.
\label{calDdef}
\end{equation}
Consider first the first term in Eq.\ (\ref{lambdadef4}).  Using the
definition (\ref{f def}) of $A_k(\alpha)$ and the definition
(\ref{intro bar sigma})
of ${\bar \sigma}_1$ and ${\bar \sigma}_2$ we can
write this term as  
\begin{equation}
{1 \over N} \sum_{k=1}^N \ln A_k(0) = - { \Delta \sigma_1^2 \over
  {\bar \sigma}_1^2}  - { \Delta \sigma_2^2 \over
  {\bar \sigma}_2^2} + O( \Delta \sigma_1^3, \Delta \sigma_2^3),
\end{equation}
where $\Delta \sigma_1 = \sigma_1 - {\bar \sigma}_1$, $\Delta \sigma_2
= \sigma_2 - {\bar \sigma}_2$.
Therefore the first term is maximized at $\sigma_1 = {\bar \sigma}_1$,
$\sigma_2 = {\bar \sigma}_2$.  Below we shall show that the second
term in Eq.\ (\ref{lambdadef4}) is of order $O(\epsilon^2)$, where in this
subsection we define $\epsilon = 1/\sqrt{N}$.  Therefore the values of
$\sigma_1$ and $\sigma_2$ that achieve the maximum are
\begin{eqnarray}
{\hat \sigma}_1 &=& {\bar \sigma}_1 \left[ 1 + O(\epsilon^2) \right] \nonumber
\\ 
{\hat \sigma}_2 &=& {\bar \sigma}_2 \left[ 1 + O(\epsilon^2) \right].
\end{eqnarray}
Moreover, in analyzing the second term it suffices to take $\sigma_1 =
{\bar \sigma}_1$, $\sigma_2 = {\bar \sigma}_2$ in order to obtain the
statistic to the leading $O(\epsilon^2)$ order.  Lastly, since we have
assumed that ${\tilde \sigma}_1 = {\tilde \sigma}_2 =1$ and no signal
is present, we have ${\bar \sigma}_{1,2} = 1 + O(\epsilon)$.  Hence,
in analyzing the second term, it is sufficient to take $\sigma_1 =
\sigma_2 = 1$.

The statistic (\ref{lambdadef4}) therefore reduces to 
\begin{equation}
{\cal L} = \max_{\alpha,\xi} 
{1 \over N} \sum_{k=1}^N \, \ln \left[ 1 + \xi {\cal D}_k(\alpha) \right] + O(\epsilon),
\label{lambdadef5}
\end{equation}
where from Eqs.\ (\ref{f def}) and (\ref{calDdef}) 
\begin{equation}
{\cal D}_k(\alpha) = {1 \over \sqrt{1 + 2 \alpha}} \exp \left[ {w_k^2
    \over 2 + {1 \over \alpha}} \right] -1.
\end{equation}
Here $w_k = (n_1^k + n_2^k)/\sqrt{2}$, $1 \le k \le N$, are
independent Gaussian random variables of zero mean and unit variance.

\begin{figure}
\begin{center}
\epsfig{file=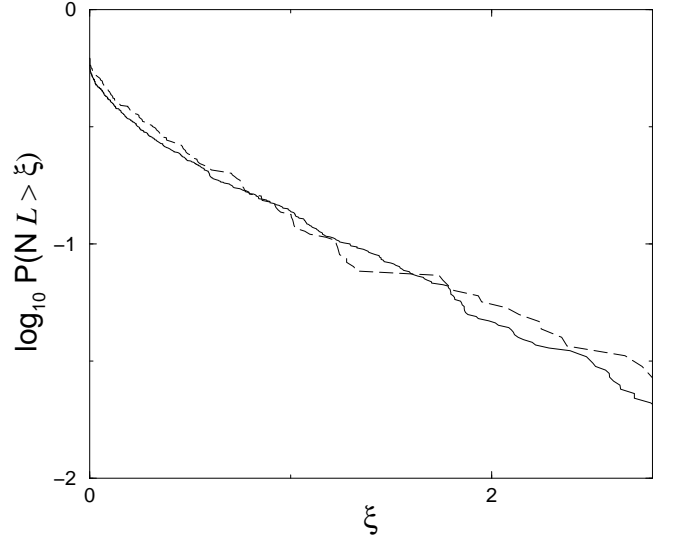,width=8.5cm}
\caption{The cumulative distribution function for the leading order
expression \protect{(\ref{lambdadef5})} for the statistic when no
signal is present, obtained numerically.  The solid line is for $N =
1000$, and the dashed line for $N = 5000$.
}
\label{fig:fa}
\end{center}
\end{figure}

It is straightforward to numerically compute the distribution of the
statistic (\ref{lambdadef5}), by generating the Gaussian variables
$w_k$ and numerically maximizing over $\xi$ and $\alpha$.  
The result is shown in Fig. \ref{fig:fa}.  We find
that at large $N$, the distribution of $N {\cal L}$ becomes
independent of $N$, and is approximately given by 
\begin{equation}
P(N {\cal L} > \xi) = \alpha_0 e^{-\beta_0 \xi}
\end{equation}
for $\xi > 0$, where $\alpha_0 \approx 0.42$ and $\beta_0 \approx
1.08$.  Therefore the false alarm probability is approximately given
by
\begin{equation}
P_{\rm FA} = \alpha_0 \exp \left[ - \beta_0 N {\cal L}_* \right].
\label{eq:ansB}
\end{equation}

Finally, we remark why it is plausible to expect the distribution of
$N {\cal L}$ to be independent of $N$ in the large $N$ limit.  The
numerical maximizations over $\xi$ and $\alpha$ in Eq.\
(\ref{lambdadef5}) show that the maximum is nearly always achieved at
$\alpha \ll 1$ or $\xi \ll 1$.  In both these regimes, one can obtain
some information about the $N$-dependence of the statistic.

Consider first the regime $\xi \ll 1$.  In this regime we can expand
the expression (\ref{lambdadef5}) as a power series in $\xi$ to obtain
\begin{equation}
{\cal L} = \max_{\alpha,\xi} 
{1 \over N} \sum_{k=1}^N \, \left[ \xi {\cal D}_k(\alpha) - {1 \over
    2} \xi^2 {\cal D}_k(\alpha)^2 + O(\xi^3) \right] + O(\epsilon).
\label{lambdadef6}
\end{equation}
The generalized central limit theorem (reviewed in Appendix
\ref{app:gclt}) implies that
\begin{equation}
{1 \over N} \sum_{k=1}^N \, {\cal D}_k(\alpha) = N^{1 - \gamma_1 \over 
  \gamma_1} \left( \ln N \right)^{\delta_1} {\cal
  F}_N(\alpha),
\label{levy1}
\end{equation}
where for each fixed $\alpha$, the distribution of the random variable ${\cal
F}_N(\alpha)$ becomes independent of $N$ in the large $N$ limit.
Here 
\begin{equation}
\gamma_1 = \left\{ \begin{array}{ll} 2 & 0 <
        \alpha \le 1/2 \\ 
        1 + {1 \over 2 \alpha} & 
        1/2 \le \alpha \\ \end{array} \right.
\end{equation}
and
\begin{equation}
\delta_1 = \left\{ \begin{array}{ll} 0 & 0 <
        \alpha \le 1/2 \\ 
         { -\alpha \over 1 + 2 \alpha} & 
        1/2 \le \alpha.\\ \end{array} \right.
\end{equation}
The limiting distribution is a Levy distribution with parameters $p =
1$ and $\gamma = \gamma_1$.
Similarly we have
\begin{equation}
{1 \over N} \sum_{k=1}^N \, {\cal D}_k(\alpha)^2 = N^{1 - \gamma_2 \over
  \gamma_2} \left( \ln N \right)^{\delta_2} {\cal 
  G}_N(\alpha),
\label{levy2}
\end{equation}
where as $N \to \infty$ at each fixed $\alpha$ the distribution of the
random variable ${\cal G}_N(\alpha)$ tends to a Levy distribution with
parameters $p=1$ and $\gamma = \gamma_2$, with
\begin{equation}
\gamma_2 = \left\{ \begin{array}{ll} 2 & 0 <
        \alpha \le 1/6 \\ 
         {1 + 2 \alpha \over 4 \alpha} & 
        1/6 \le \alpha.\\ \end{array} \right.
\end{equation}
and 
\begin{equation}
\delta_2 = \left\{ \begin{array}{ll} 0 & 0 <
        \alpha \le 1/6 \\ 
         { - 2 \alpha \over 1 + 2 \alpha} & 
        1/6 \le \alpha. \\ \end{array} \right.
\end{equation}

We now substitute the results (\ref{levy1}) and (\ref{levy2}) into the
expression (\ref{lambdadef6}) for the statistic, and maximize analytically over
the quadratic dependence on $\xi$.  For $\alpha \ge 1/2$, the value of
$\xi$ which achieves the maximum goes to zero as $N \to \infty$,
consistent with the assumption $\xi \ll 1$, and the result is
\footnote{For $\alpha < 1/2$ this argument fails, which is why we must
numerically verify that the distribution of $N {\cal L}$ is
asymptotically independent of $N$.}
\begin{equation}
N {\cal L} = {1 \over 2 } \max_\alpha {{\cal F}_N(\alpha)^2 \over {\cal
    G}_N(\alpha)} + O(\epsilon).
\end{equation}

In the regime $\alpha \ll 1$, if we expand the expression
(\ref{lambdadef5}) to quadratic order in $\alpha$, the result is an
expression which is a linear function of $1/\xi$ at fixed $\alpha
\xi$.  Hence, when one 
maximizes over values of $\xi$ in the range $0 \le \xi \le 1$, the
maximum is always achieved either at $\xi =0$ or $\xi =1$.  One can
show that the maximum to this order is always achieved at $\xi = 1$, and the
resulting expression is
\begin{equation}
N {\cal L} = {1 \over 4 } {\cal G}^2 + O(\epsilon),
\end{equation}
where 
\begin{equation}
{\cal G} = \sqrt{N} \left[ {1 \over N} \sum_{k=1}^N w_k^2 \ -1 \right]
\end{equation}
has a distribution that is independent of $N$ in the large $N$ limit.

\section{Generalized central limit theorem}
\label{app:gclt}

In this appendix we review the generalized central limit theorem that
can be found on p.~574 of Ref.~\cite{Feller}.  First we define a
particular distribution function called the Levy distribution.  It
depends on 3 real parameters, a positive constant $C$, a parameter
$\gamma$ in the range $0 < \gamma \le 2$, and constant $p$ in the
range $0 \le p \le 1$ \footnote{The parameter $\gamma$ is conventionally denoted by $\alpha$.  We use $\gamma$ here to avoid confusion with the variable
$\alpha$ defined in Eq.\ (\ref{eq:sigg}).}.  We say a random 
variable $X$ has a Levy distribution with parameters $C$, $\gamma$ and
$p$ if the characteristic function of $X$ is given by
\begin{eqnarray}
\left< e^{i \zeta X} \right> &=& \exp \bigg\{ | \zeta|^\gamma { C
    \Gamma(3-\gamma) \over \gamma (\gamma-1)} \bigg[ \cos(\pi
    \gamma/2) \nonumber \\
  &&  + i \, {\rm sgn}(\zeta) (p-q) \sin(\pi \gamma/2) \bigg] \bigg\},
\end{eqnarray}
where $q = 1 - p$.
The corresponding probability distribution function  is obtained by
taking a Fourier transform and decays like $x^{-(1+\gamma)}$ at large
$x$ for $\gamma < 2$ ($\gamma =2$ is the Gaussian case).

Consider now a random
variable $X$ with probability
distribution function $f(x)$ whose variance is infinite.  Let
\begin{equation}
F(x) = \int_{-\infty}^x dy \, f(y)
\end{equation}
be the cumulative distribution function
and define 
\begin{equation}
\mu(x) = \int_{-x}^x dy y^2 f(y).
\end{equation}
Suppose that
the distribution satisfies the following conditions:
(i)  As $x \to \infty$ we have $\mu(x) \sim x^{2 - \gamma} L(x)$,
where $0 < \gamma \le 2$, and $L(x)$ varies slowly in the sense that
$L(tx)/L(t) \to 1$ as $t \to \infty$ for all $x>0$.  (ii) We have
\begin{equation}
{1 - F(x) \over F(-x) + 1 - F(x)} \to p \ \ \ \ \ {F(-x) \over F(-x) +
  1 - F(x)} \to q
\end{equation}
as $x \to \infty$, where $0 \le p \le 1$, $0 \le q \le 1$ and $p+q=1$.
(iii) For $1 < \gamma \le 2$, we assume that the expected value $\int
dx \, x f(x)$ vanishes; this can be enforced by making a
transformation of the form $X \to X + {\rm constant}$.

We define the sequence of random variables
\begin{equation}
S_N = {1 \over a_N} \sum_{i=1}^N \, X_i,
\end{equation}
where the $X_i$ are independent, identically distributed random
variables with distribution function $f$, and the constants
$a_N$ are 
chosen to satisfy
\begin{equation}
{N \mu(a_N)  \over a_N^2} \to C
\end{equation}
as $N \to \infty$, where $C$ is a positive constant.  Then, the
distribution functions of the random
variables $S_N$ converge to a Levy distribution with parameters $C$,
$\gamma$ and $p$ as $N \to \infty$.

\end{document}